%% file: paperDE_arxiv.tex
\documentclass[11pt,a4paper]{article}
\usepackage{graphicx}
\usepackage{natbib}
\usepackage{color}
\usepackage{amsmath}
\usepackage{url}
\usepackage{bm}
\usepackage{bibunits}

\newenvironment{methods}{}{}
\defaultbibliographystyle{natbib}
\defaultbibliography{paperDE}


\def\textcolon{\text{\rm :}}

\usepackage{graphicx}
\usepackage[tight]{subfigure}
\usepackage{amsmath}
\usepackage{amssymb}
\usepackage{amsthm}
\usepackage{url}

\newcommand{\TS}{\theta^{(*)}}
\newcommand{\TA}{\theta^{act}}
\newcommand{\ZA}{Z_n^{act}}
\newcommand{\Brack}[1]{\left( #1 \right) }
\newcommand{\COMMENT}[1]{}

\begin{document}
\hyphenation{tran-script-ome}

\title{Identifying differentially expressed transcripts from RNA-seq data with biological variation}
\author{Peter Glaus\,$^{1,}$\footnote{to whom correspondence should be addressed}, Antti Honkela\,$^{2,\ast}$\footnote{These authors contributed equally to this work.} and Magnus Rattray\,$^{3,\ast,\dagger}$ \\[5mm]
\small $^{1}$School of Computer Science, University of Manchester,
\small Oxford Road, Manchester M13 9PL, UK\\
\small $^{2}$Helsinki Institute for Information Technology HIIT,
\small Department of Computer Science, \\
\small University of Helsinki, Helsinki, Finland \\
\small $^{3}$Department of Computer Science and 
\small Sheffield Institute of Translational Neuroscience,\\
\small The University of Sheffield, Sheffield S10 2HQ, UK
}

\maketitle

\begin{bibunit}

\input{abstract_arxiv}

\input{introduction}

\input{methods}

\input{results}

\input{conclusion}

\section*{Acknowledgement}
\paragraph{Funding\textcolon}
This work was supported under the European ERASysBio+ initiative
project SYNERGY by the Biotechnology and Biological Sciences Research
Council [BB/I004769/2 to M.R.] and the Academy of Finland [135311 to
A.H.]; by the Academy of Finland [121179 to A.H.]; and
the IST Programme of the European Community, under the PASCAL2
Network of Excellence [IST-2007-216886].
This publication only reflects the authors' views.

\small

\putbib

\end{bibunit}

\appendix
\clearpage
\begin{bibunit}
\section*{Supplementary Information}

\input{supplement_body}

\putbib
\end{bibunit}

\end{document}

%% file: abstract_arxiv.tex
\begin{abstract}

\textbf{Motivation:}
High-throughput sequencing enables expression analysis at the level of individual transcripts. 
The analysis of transcriptome expression levels and differential expression estimation requires a probabilistic approach to properly account for ambiguity caused by shared exons and finite read sampling as well as the intrinsic biological variance of transcript expression.

\textbf{Results:}
We present BitSeq (Bayesian Inference of Transcripts from Sequencing data), a Bayesian approach for estimation of transcript expression level from RNA-seq experiments. 
Inferred relative expression is represented by Markov chain Monte Carlo (MCMC) samples from the posterior probability distribution of a generative model of the read data.
We propose a novel method for differential expression analysis across replicates which propagates uncertainty from the sample-level model while modelling biological variance using an expression-level-dependent prior.
We demonstrate the advantages of our method using simulated data as well as an RNA-seq dataset with technical and biological replication for both studied conditions.

\textbf{Availability:}
The implementation of the transcriptome expression estimation and differential expression analysis, BitSeq, has been written in \texttt{C++}.

\textbf{Contact:} \url{glaus@cs.man.ac.uk}, \url{antti.honkela@hiit.fi}, \url{M.Rattray@sheffield.ac.uk}
\end{abstract}


%% file: introduction.tex
\section{Introduction}

High-throughput sequencing is an effective approach for transcriptome analysis.
This methodology, also called RNA-seq, has been used to analyse unknown transcript sequences, estimate gene expression levels and study single nucleotide polymorphisms \citep{Wang2009}. As shown by other researchers \citep{Mortazavi2008}, RNA-seq provides many advantages over microarray technology, although effective analysis of RNA-seq data remains a challenge.

A fundamental task in the analysis of RNA-seq data is the identification of a set of differentially expressed genes or transcripts. Results from a differential expression (DE) analysis of individual transcripts are essential in a diverse range of problems such as identifying differences between tissues~\citep{Mortazavi2008}, understanding developmental changes~\citep{Graveley2011} and regulator such as microRNA target prediction \citep{Xu2010a}. To carry out an effective DE analysis it is important to obtain accurate estimates of expression for each sample but it is equally important to properly account for all sources of variation, technical and biological, to avoid spurious DE calls \citep{Robinson2007,Anders2010,Oshlack2010}. In this contribution we address both of these problems by developing integrated probabilistic models of the read generation process and the biological replication process in an RNA-seq experiment.

During the RNA-seq experimental procedure, a studied specimen of transcriptome is synthesised into cDNA, amplified, fragmented and then sequenced by a high-throughput sequencing device. This process results in a dataset consisting of up to hundreds of millions of short sequences, or reads, encoding observed nucleotide sequences. 
The length of the reads depends on the sequencing platform and currently typically ranges from 25 to 300 base pairs. 
Reads have to be either assembled into transcript sequences or aligned to a reference genome by an aligning tool, to determine the sequence they originate from.

With proper sample preparation, the number of reads aligning to a certain gene is approximately proportional to the abundance of fragments of transcripts for that gene within the sample \citep{Mortazavi2008} allowing researchers to study gene expression \citep{Cloonan2008,Marioni2008}.
However, during the process of transcription, most eukaryotic genes can be spliced into different transcripts which share parts of their sequence. 
As it is the transcripts of genes that are being sequenced during RNA-seq, it is possible to distinguish between individual transcripts of a gene. Several methods have been proposed to estimate transcript expression levels \citep{Li2010,Nicolae2010,Katz2010,Turro2011}. Furthermore, \citet{Wang2010c} showed that estimating gene expression as a sum of transcript expression levels yields more precise results than inferring the gene expression by summing reads over all exons.

\begin{figure*}[!tpb]
   \centering
   \includegraphics[width=0.72\textwidth]{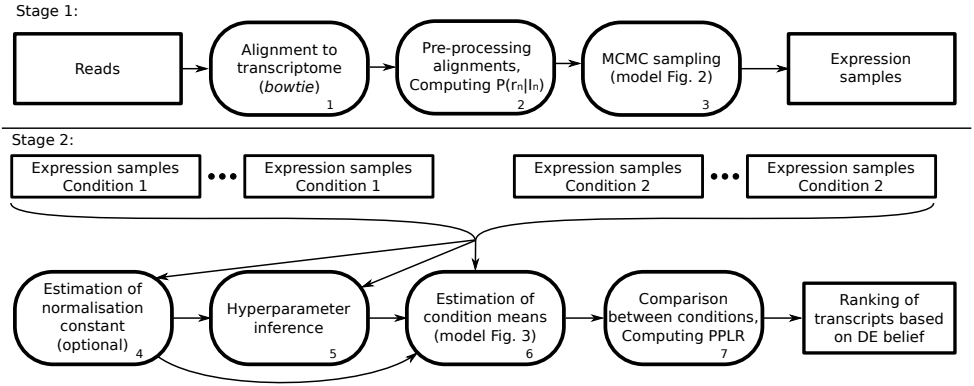}
   \caption{Diagram showing the BitSeq analysis pipeline divided into two separate stages. In Stage 1, transcript expression levels are estimated using reads from individual sequencing experiments. In step 1, reads are aligned to the transcriptome. In step 2, the probability of a read originating from a given transcript $P(r_n|I_n)$ is computed for each alignment based on Eq.~(\ref{eq:alignmentProbability}). These probabilities are used in step 3 of the analysis, MCMC sampling from the posterior distribution in Eq.~(\ref{eq:posterior}).
   In Stage 2 of the analysis, the posterior distributions of transcript expression levels from multiple conditions and replicas are used to infer the probability that transcripts are differentially expressed. In step 4, a suitable normalisation for each experiment is estimated. The normalised expression samples are further used to infer expression-dependent variance hyperparameters in step 5. Using these results, replicates are summarized by estimating the per-condition mean expression for each transcript, Eq.~(\ref{eq:modelDE}), in step 6. Finally, in step 7, samples representing the distribution of within-condition expression are used to estimate the probability of positive log ratio (PPLR) between conditions, which is used to rank transcripts based on DE belief.
   }\label{fig:pipeline}
\end{figure*}

Since the transcript of origin is undecidable for reads aligning to shared subsequence, estimation of transcript expression levels has to be completed in a probabilistic manner. 
Initial studies of transcript expression used the Expectation-Maximization (EM) approach \citep{Li2010,Nicolae2010}. This is a maximum likelihood procedure which only provides a point estimate of transcript abundance and does not measure the uncertainty in these estimates. 
To overcome this limitation, \citet{Katz2010} used a Bayesian approach to capture the posterior distribution of the transcript expression levels using a Markov chain Monte Carlo (MCMC) algorithm.
\citet{Turro2011} have also proposed MCMC estimation for a model of read counts over regions that can correspond to exons or other suitable subparts of transcripts.

In this contribution we present BitSeq (Bayesian Inference of Transcripts from Sequencing data), a new method for inferring transcript expression and analysing expression changes between conditions.
We use a probabilistic model of the read generation process similar to the model of \citet{Li2010} and we develop an MCMC algorithm for Bayesian inference over the model. 
\citet{Katz2010} developed an MCMC algorithm for a similar generative model but our model differs from theirs because we allow for multi-aligned reads mapping to different genes. 
Furthermore, we infer the overall relative expression of transcripts across the transcriptome whereas \citet{Katz2010} focus on relative expression of transcripts from the same gene. 
We have implemented MCMC using a collapsed Gibbs sampler to sample from the posterior distribution of model parameters. 

In many gene expression studies expression levels are used to select genes with differences in expression in two conditions, a process referred to as a DE analysis. 
We propose a novel method for DE analysis that includes a model of biological variance while also allowing for the technical uncertainty of transcript expression which is represented by samples from the posterior probability distribution obtained from the probabilistic model of read generation. 
By retaining the full posterior distribution, rather than a point estimate summary, we can propagate uncertainty from the initial read summarization stage of analysis into the DE analysis. 
Similar strategies have been shown to be effective in the DE analysis of microarray data \citep{Liu2006,Rattray2006} but given the inherent uncertainty of reads mapping to multiple transcripts we expect the approach to bring even more advantages for transcript-level DE analyses.
Furthermore, this method accounts for decreased technical
reproducibility of RNA-seq for low-expressed transcripts recently
reported by \citet{Labaj2011} and can decrease the number of
transcripts falsely identified as differentially expressed.


%% file: methods.tex
\begin{methods}

\section{Methods}

The BitSeq analysis pipeline consists of two main stages: transcript expression estimation and differential expression assessment, see Figure \ref{fig:pipeline}. 
For the transcript expression estimation the input data are single-end or pair-end reads from a single sequencing run. The method produces samples from the inferred probability distribution over transcripts' expression levels.
This distribution can be summarized by the sample mean in case one is only interested in expression.

The DE analysis uses posterior samples of expression levels from two or more conditions and all available replicates.
The conditions are summarized by inferring the posterior distribution of condition mean expression. Samples from the 
posterior distributions are compared to score the transcripts based on the belief in change of 
expression level between conditions. 


\subsection{Stage 1: Transcript expression estimation}

The initial interest when dealing with RNA-seq data is estimation of expression levels within a sample. 
In this work, we focus on the transcript expression levels, mainly represented by $\boldsymbol{\theta}=(\theta_1,\dots ,\theta_M)$, the relative abundance of transcripts' fragments within the studied sample, where $M$ is the total number of transcripts.
This can be further transformed into relative expression of transcripts $\TS_m = \theta_m / ( l_m (\sum_{i=1}^M \theta_i /l_i))$, where $l_m$ is the length of the $m$-th transcript. 
Alternatively, expression can be represented by \textit{reads per kilobase per million mapped reads}, $RPKM_m = \theta_m \times 10^9 / l_m$, introduced by \citet{Mortazavi2008}.

\begin{figure}[t] 
   \centering
   \includegraphics[scale=0.36]{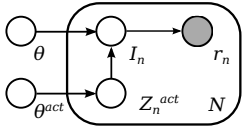}
   \caption{
      Graphical representation of the RNA-seq data probabilistic model. 
      We can consider the observation of reads $R=(r_1,\dots,r_N)$ as $N$ conditionally independent events, with each observation of a read $r_n$ depending on the transcript (or isoform) it originated from $I_n$.
      The probability of sequencing a given transcript $I_n$ depends on the relative expression of fragments $\boldsymbol{\theta}$ and the noise indicator $\ZA$.
      The noise indicator variable $\ZA$ depends on noise parameter $\TA$, and indicates that the transcript being sequenced is regarded as noise, which enables observation of low quality and un-mappable reads.
   }\label{fig:modelE}
\end{figure} 

We use a generative model of the data, depicted in Figure \ref{fig:modelE}, which models the RNA-seq data as independent observations of individual reads $r_n\in R=\{r_1,\dots,r_N\}$, depending on the relative abundance of transcripts' fragments $\boldsymbol{\theta}$ and a noise parameter $\TA$. 
The parameter $\TA$ determines the number of reads regarded as noise and enables the model to account for unmapped reads as well as for low-quality reads within a sample.

Based on the parameter $\TA$, indicator variable $\ZA\sim \textrm{Bern}(\TA)$ determines whether read $r_n$ is considered as noise or a valid sequence. 
For a valid sequence, the process of sequencing is being modelled.
Under the assumption of reads being uniformly sequenced from the molecule fragments, each read is assigned to a transcript of origin by the indicator variable $I_n$, which is given by categorical distribution $I_n\sim \textrm{Cat}(\boldsymbol{\theta})$.

For a transcript $m$ we can express the probability of an observed alignment as the probability of choosing a specific position $p$ and sequencing a sequence of given length with all its mismatches, $P(r_n|I_n=m) = P(p|m)P(r_n|seq_{mp})$.
For paired-end reads we compute the joint probability of the alignment of a whole pair, in which case we also have to consider fragment length distribution $P(l)$,
\begin{multline}\label{eq:alignmentProbability}
P(r_n^{(1)},r_n^{(2)}|I_n=m) = \\
P(p|l,m)P(l|m)P(r_n^{(1)}|seq_{mlp_1})P(r_n^{(2)}|seq_{mlp_2}) \ .
\end{multline}
Details of alignment probability computation including optional
position and sequence-specific bias
correction methods are presented in Supplementary Material.
For every aligned read, we also calculate the probability that the read is from neither of the aligned transcripts, but is regarded as sequencing error or noise $P(r_n|\textrm{noise})$. 
This value is calculated by taking the probability of the least probable valid alignment corrupted with two extra base mismatches.

The joint probability distribution of the model can now be written as 
\begin{equation}
\begin{array}{rl}
   &P(R, \boldsymbol{I}, \boldsymbol{Z^{act}}, \boldsymbol{\theta}, \TA) =  P(\boldsymbol{\theta})P(\TA) \\
   & \times \prod_{n=1}^N \Brack{P(r_n|I_n)P(I_n|\boldsymbol{\theta},\ZA)P(\ZA|\TA)},
\end{array}
\end{equation}
where we use weak conjugate Dirichlet and Beta prior distributions for $\boldsymbol{\theta}$ and $\TA$, respectively. 
The posterior distribution of the model's parameters given the data $R$ can be simplified by integrating over all possible values of  $Z^{act}$:
\begin{equation}
\begin{array}{rl}
   P(\boldsymbol{I},\boldsymbol{\theta},\TA|R) 
   & \!\!\!\! \propto  P(\boldsymbol{\theta})P(\TA) 
   \prod_{n;I_{n}\neq 0} \Brack{P(r_n|I_n) \textrm{Cat}(I_n|\boldsymbol{\theta})\TA} \\
   & \times\prod_{n;I_{n}=0} \Brack{P(r_n|\textrm{noise})(1-\TA)} \label{eq:posterior}\ .
\end{array}
\end{equation}

According to the model any read can be a result of sequencing either strand of an arbitrary transcript at a random position.
However, the probability of a read originating from a location where it does not align is negligible.
Thus the term $P(r_n|I_n) \textrm{Cat}(I_n|\boldsymbol{\theta)}\TA$ has to be evaluated only for transcripts and positions to which the read does align.
To accomplish this we first align the reads to the
transcript sequences using the \textit{Bowtie} alignment tool
\citep{Langmead2009}, preserving possible multiple alignments to
different transcripts.
We then pre-compute $P(r_n|I_n)$ only for the valid alignments.
(See steps 1-2 in Figure~\ref{fig:pipeline}.)

The closed form of the posterior distribution is not analytically tractable and an approximation has to be used.
We can analytically marginalise $\boldsymbol{\theta}$ and apply a
collapsed Gibbs sampler to produce samples from the posterior
probability distribution over $I_n$~\citep{Geman1993,Griffiths2004}.
These are used to compute a posterior for $\boldsymbol{\theta}$, which
is the main variable of interest.
Full update equations
for the sampler are given in Supplementary Material.

In the MCMC approach, multiple chains are sampled at the same time and convergence is monitored using the $\widehat{R}$ statistic as described by \citet{bda}.
The $\widehat{R}$ statistic is an estimate of a possible scale reduction of the marginal posterior variance and provides a measure of usefulness of producing more samples.
Posterior samples of $\boldsymbol{\theta}$ provide an assessment of the abundance of individual transcripts. As well as providing an accurate point estimate of the expression levels through the mean of the posterior, the probability distribution provides a measure of confidence for the results, which can be used in further analyses.

\subsection{Stage 2: Combining data from multiple replicates
  and estimating differential expression}

To identify transcripts that are truly differentially expressed it is necessary to account for biological variation by using replication for each experimental condition.
Our method summarizes these replicates by estimating the biological variance and inferring per-condition mean expression levels for each transcript.
During the differential expression analysis we consider the logarithm of transcript expression levels $y_m = \log \theta_m$.
The model for data originating from multiple replicates is illustrated in Figure \ref{fig:modelDE}. We use a hierarchical
log-normal model of within-condition expression. The prior over the biological variance is dependent on the mean expression level
across conditions and the prior parameters (hyper-parameters) are learned from all of the data by fitting a non-parametric
regression model. We fit a model 
for each gene using the expression estimates from Stage 1. 

A novel aspect of our Stage 2 approach is that we fit 
models to posterior samples obtained from the MCMC simulation from Stage 1, which can be considered 
``pseudo-data'' representing expression corrupted by technical noise. A pseudo-data vector 
in constructed using a single MCMC sample for each
replicate across all conditions. The posterior distribution over per-condition means is inferred for each
pseudo-data vector using the model in Figure~\ref{fig:modelDE} (described below). We then use Bayesian model-averaging 
to combine the evidence from each pseudo-data vector and determine the probability
of differential expression. This approach allows us to account for the intrinsic technical
variance in the data; it is also computationally tractable because the model for a single pseudo-data vector is conjugate 
and therefore inference can be carried out exactly. This effectively regularizes 
our variance estimate in the case that the number of replicates is low. As shown in Section~\ref{sec:de_comparison} this 
provides improved control of error rates for 
weakly expressed transcripts where the technical 
variance is large.

\begin{figure}[!tpb]
   \centering
   \includegraphics[scale=0.36]{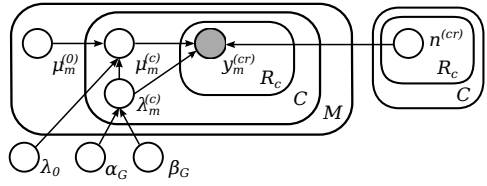}
   \caption{Graphical model of the biological variance in transcript expression experiment. 
      For replicate $r$, condition $c$ and transcript $m$, the observed log-expression level $y_m^{(cr)}$ is normally distributed around the normalised condition mean expression $\mu_m^{(c)}+n^{(cr)}$ with biological variance $1/\lambda_m^{(c)}$.
      The condition mean expression $\mu_m^{(c)}$ for each condition is normally distributed with overall mean expression $\mu_m^{(0)}$ and scaled variance $1/(\lambda^{(c)}_m\lambda_0)$.
      The inverse variance, or precision $\lambda^{(c)}_m$, for a given transcript $m$ follows a Gamma distribution with expression-dependent hyperparameters $\alpha_G,\beta_G$, which are constant for a group of transcripts $G$ with similar expression.
   }\label{fig:modelDE}
\end{figure}

For a condition $c$ we assume $R_c$ replicate datasets. 
The log-expression from replicate $r$, $y_m^{(cr)}$ is assumed to be distributed according to a normal distribution with condition mean expression $\mu_m^{(c)}$, normalised by replication specific constant $n^{(cr)}$, and precision $\lambda_m^{(c)}$, $y_m^{(cr)}\sim \textrm{Norm}(\mu_m^{(c)}+n^{(cr)},1/\lambda_m^{(c)})$.
As our parameters represent the relative expression levels in the sample, 
BitSeq implicitly incorporates normalisation by the total number of reads or the RPKM measure, as was done when generating the results in this publication.
Further normalisation can be implemented using
the normalisation constant $n^{(cr)}$, which is constant for all transcripts of a given replicate and can be estimated prior to probabilistic modeling using, for example, a quantile based method~\citep{Robinson2010} or any other suitable technique.

The condition mean expression is normally distributed $\mu_m^{(c)}\sim \textrm{Norm}(\mu_m^{(0)},1/(\lambda_m^{(c)}\lambda_0))$ with mean $\mu_m^{(0)}$, which is empirically calculated from multiple samples, and scaled precision $\lambda_m^{(c)}\lambda_0$.
The prior distribution over per-transcript, condition specific precision $\lambda_m^{(c)}$ is a Gamma distribution with hyperparameters $\alpha_G,\beta_G$, which are fixed for a group of transcripts with similar expression level, $G$.

The hyperparameters $\alpha_G,\beta_G$ determine the distribution over per-transcript precision parameter $\boldsymbol{\lambda_m}$ which varies with the expression level of a transcript (see Supplementary Figure~3 of the supplementary material).
For this reason, we inferred these hyperparameters from the dataset for various levels of expression, prior to the estimation of precision $\boldsymbol{\lambda_m}$ and mean expression $\boldsymbol{\mu_m}$.
We used the same model as Figure \ref{fig:modelDE} applied jointly to multiple transcripts with similar empirical mean expression levels $\mu_m^{(0)}$. 
We set a uniform prior for the hyperparameters, marginalized out condition means and precision, and used a MCMC algorithm to sample $\alpha_G,\beta_G$.
The samples of $\alpha_G,\beta_G$ were smoothed by Lowess regression \citep{Cleveland1981} against empirical mean expression to produce a single pair of hyperparameters for each group of transcripts with similar expression level.

This model is conjugate and thus leads to a closed form posterior distribution. This allows us to directly sample $\boldsymbol{\lambda_m}$ and $\boldsymbol{\mu_m}$ given each pseudo-data vector $\boldsymbol{y_m}$ constructed from the Stage 1 MCMC samples:
\begin{align} \label{eq:modelDE}
& P(\boldsymbol{\mu_m},\boldsymbol{\lambda_m}|\boldsymbol{y_m})  = {\textstyle \prod_{c=1}^C} \textrm{Gamma}(\lambda_m^{(c)}|a_c,1/b_c) \notag\\
 &  \textrm{Norm}\Brack{\mu_m^{(c)}\left|{\textstyle \frac{\mu_m^{(0)}\lambda_0 + \sum_{r=1}^{R_c} (y_m^{(cr)}-n^{(cr)})}{\lambda_0 + R_c}, \frac{1}{\lambda_m^{(c)}(\lambda_0+R_c)}}\right.} ,\\
& a_c = \alpha_G + {\textstyle \frac{R_c}{2} }, \notag\\
& b_c = \beta_G + {\textstyle \frac{1}{2} }\left((\mu_m^{(0)})^2\lambda_0 + \right. \notag\\
& + \left.
\textstyle \sum_{r=1}^{R_c} \left(y_m^{(cr)}-n^{(cr)}\right)^2 - \frac{\Brack{\mu_m^{(0)}\lambda_0 + \sum_{r=1}^{R_c} \left(y_m^{(cr)}-n^{(cr)}\right)}^2}{\lambda_0+R_c}
\right). \notag
\end{align}


Samples of $\mu_m^{(c_1)}$ and $\mu_m^{(c_2)}$ are used to compute the probability of expression level of transcript $m$ in condition $c_1$ being greater than the expression level in condition $c_2$. This is done by counting the fraction of samples in which the mean expression from the first condition is greater, that is $P(\mu_m^{(c_1)}>\mu_m^{(c_2)}|R) = \frac{1}{N}\sum_{n=1}^{N} \delta(\mu_{m,n}^{(c_1)} > \mu_{m,n}^{(c_2)})$ which we refer to as the Probability of Positive Log-Ratio (PPLR). Here, $n=1\ldots N$ represents one sample from the above posterior distribution for each of $N$ independent pseudo-data vectors.
Subsequently, ordering transcripts based on PPLR produces a ranking of most probable up-regulated and down-regulated transcripts. This kind of one-sided Bayesian test has previously been used for the analysis of microarray data \citep{Liu2006}.

\end{methods}

%% file: results.tex
\section{Results and Discussion}

\subsection{Datasets}

We carried out experiments evaluating both gene expression estimation accuracy as well as differential expression analysis precision. 
For the evaluation of bias correction effects as well as comparison with other methods (Table \ref{tab:exp}) we used paired-end RNA-seq data from the Microarray Quality Control (MAQC) project \citep{Shi2006} (Short Read Archive accession number SRA012427), because it contains 907 transcripts which were also analysed by TaqMan qRT-PCR. 
The results from qRT-PCR probes are generally regarded as ground truth expression estimates for comparison of RNA-seq analysis methods \citep{Roberts2011}.
We used RefSeq refGene transcriptome annotation, assembly NCBI36/hg18 in order to keep results consistent with qRT-PCR data as well as previously published comparisons by \citet{Roberts2011}.

The second dataset used in our evaluation was originally published by \citet{Xu2010a} in a 
study focused on identification of microRNA targets and provides technical as well as biological replicates for both studied conditions. 
We use this data to illustrate the importance of biological replicates for DE analysis (Figure \ref{fig:de_dist}, Supplementary Figure 3 for biological variance) and the advantages of using a Bayesian approach for both expression inference and DE analysis (Figure \ref{fig:cor_hist}).

For the purpose of evaluating and comparing BitSeq to existing differential expression analysis methods, we created artificial RNA-seq datasets with known expression levels and differentially expressed transcripts.
We selected all transcripts of chromosome 1 from human genome assembly NCBI37/hg19 and simulated two biological replicates for each of the two conditions. 
We initially sample the expression for all replicates using the same mean relative expression and variation between replicates as were observed in the Xu et al. data estimates. Afterwards we randomly choose one third of the transcripts and shift one of the conditions up or down by a known fold change.
Given the adjusted expression levels, we generated 300k single-end reads uniformly distributed along the transcripts.
The reads were reported in Fastq format with Phred scores randomly generated according to empirical distribution learned from the SRA012427 dataset.
With the error probability given by a Phred score, we generated base mismatches along the reads.

\subsection{Expression level inference}

\begin{figure}[!tpb]
   \centering
   \subfigcapskip-0.5ex
   \subfigcapmargin1ex
   \subfigure[Anti-correlation of transcripts.]{
   \includegraphics[width=0.21\textwidth]{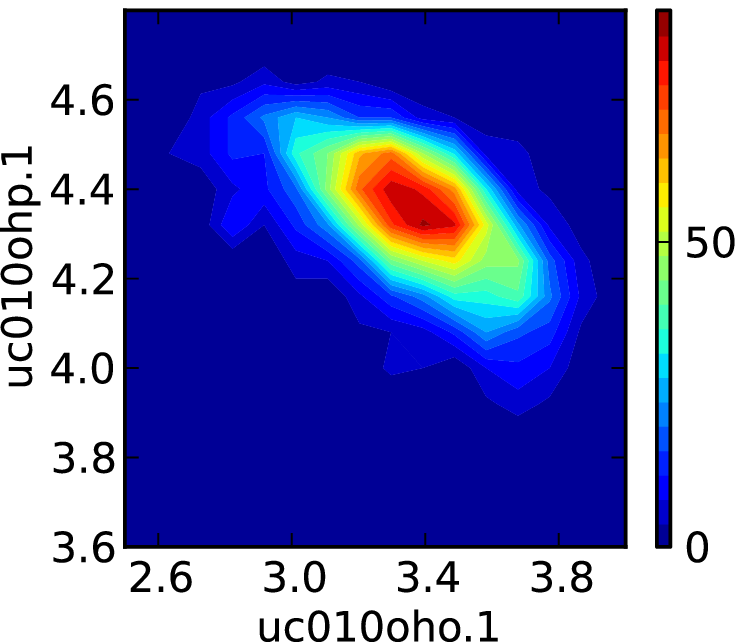}
   }%
   \subfigure[No observable correlation.]{
   \includegraphics[width=0.21\textwidth]{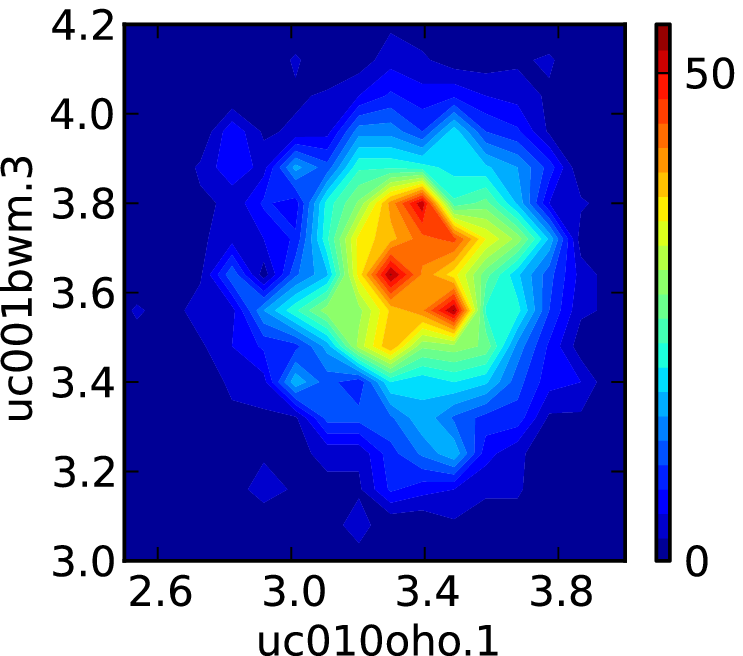}
   }
   \subfigure[Posterior distribution of expression levels for each transcript.] {
   \includegraphics[width=0.42\textwidth]{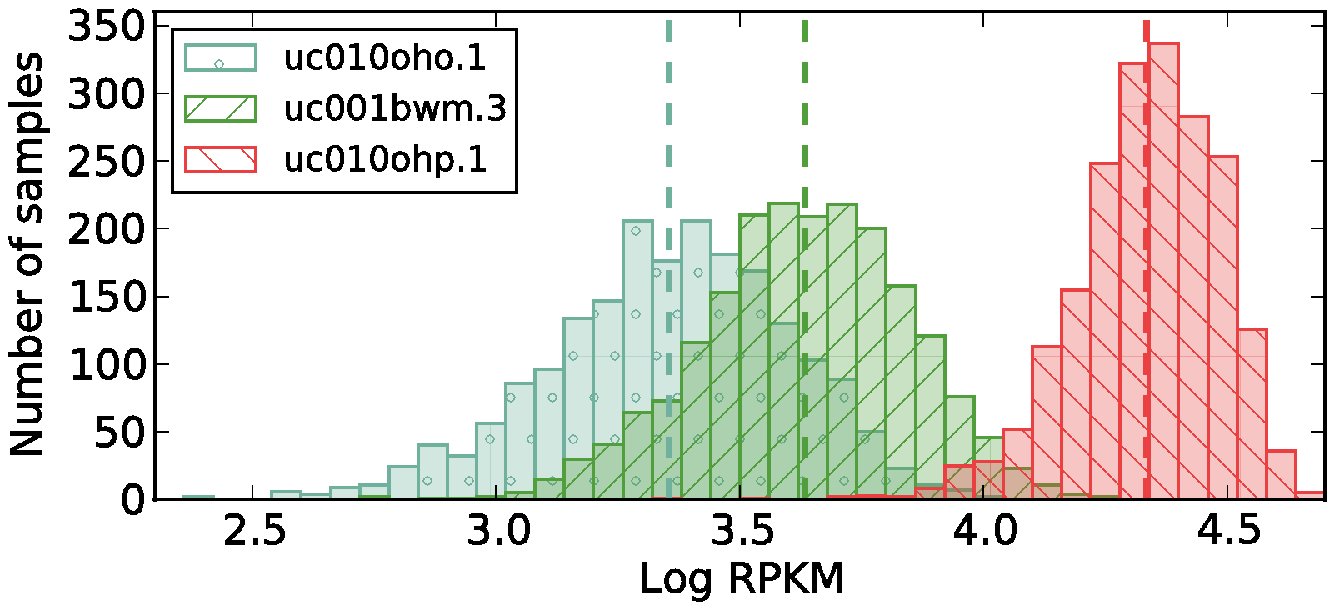}
   }
   \caption{In plots (a) and (b) we show the posterior transcript expression density for pairs of transcripts from the same gene. This is a density map constructed using the MCMC expression samples for these three transcripts. In (c) we show the marginal posterior distribution of expression levels of the same transcripts as illustrated by histograms of MCMC samples. 
   The sequencing data is from miRNA-155 study published by \citet{Xu2010a}.
}\label{fig:cor_hist}
\end{figure}

Figure \ref{fig:cor_hist} demonstrates the ambiguity that may be present in the process of expression estimation.
In Figures \ref{fig:cor_hist}(a) and \ref{fig:cor_hist}(b) we show the density of samples from the posterior distribution of expression levels for two pairs of transcripts. 
The expression levels of transcripts uc010oho.1 and uc010ohp.1 (Fig. \ref{fig:cor_hist}(a)) are negatively correlated.
On the other hand transcripts uc010oho.1 and uc001bwm.3 exhibit no visible correlation (Fig. \ref{fig:cor_hist}(b)) in their expression level estimates.
Even though this kind of correlation does not have to imply biological significance, it does point to technical difficulties in the estimation process.
These transcripts share a significant amount of sequence and the consequent read mapping ambiguity leads to greater uncertainty in expression estimates (See Supplementary Figure 1(d) for transcript profile). Bayesian inference can be used to assess the uncertainty due to such confounding factors, unlike the maximum likelihood point estimates provided by an EM algorithm. The marginal posterior probability of transcript expression for each transcript is shown in Figure \ref{fig:cor_hist}(c). 
In our analysis pipeline, the marginal posterior distributions are propagated into the differential expression estimation stage, thus the uncertainty from expression estimation is taken into account when assessing whether there is strong evidence that transcripts are differentially expressed.

\subsection{Expression estimation accuracy and read distribution bias correction}

\begin{table}[!tpb]
   \centering
\COMMENT{\begin{tabular}{l|c|c}
method & uniform & non-uniform \\
\hline
BitSeq    & \textbf{0.7677} & 0.8011 \\
Cufflinks & 0.7503 & \textbf{0.8056} \\
RSEM      & 0.7632 & 0.7633 \\
MMSEQ     & 0.7614 & --- 
\end{tabular}}
\begin{tabular}{|c|c|c|c|c|}
\hline
read model  & BitSeq & Cufflinks & RSEM & MMSEQ \\
\hline 
uniform     & \textbf{0.7677} & 0.7503 & 0.7632 & 0.7614 \\
non-uniform & 0.8011 & \textbf{0.8056} & 0.7633 & ---  \\
\hline
\end{tabular}

\bigskip

\caption{Comparison of expression estimation accuracy against TaqMan qRT-PCR data and the effect of non-uniform read distribution models using correlation coefficient $R^2$ of average expression from three technical replicates with the 893 matching transcripts analysed by qRT-PCR. 
The sequencing data (SRA012427) is part of the MAQC project and was originally published by \citet{Shi2006}.
}\label{tab:exp}
\end{table}

Initially, it was assumed that high-throughput sequencing produces reads uniformly distributed along transcripts. 
However, more recent studies show biases in the read distribution depending on the position and surrounding sequence~\citep{Dohm2008,Wu2011,Roberts2011}.
Our generative model for transcript expression inference (Figure \ref{fig:modelE}) includes a model of the underlying read distribution which in the $P(r_n|I_n=m)$ term that is calculated as a pre-processing step.
The current BitSeq implementation contains the option of using a uniform read density model or using the model proposed by \citet{Roberts2011} which
can account for positional and sequence bias. 
The effect of correcting for read distribution was analysed using the SRA012427 dataset and results are presented in Table \ref{tab:exp}.
We also compare BitSeq with three other transcript expression estimation methods: Cufflinks v0.9.3 \citep{Roberts2011}, MMSEQ v0.9.18 \citep{Turro2011} and RSEM v1.1.14 \citep{Li2011}. 

The dataset contains three technical replicates. These were analysed separately and the resulting estimates for each method were averaged together.
Subsequently, we calculated the squared Pearson correlation coefficient ($R^2$) of the average expression estimate and the results of qRT-PCR analysis.
All four methods used with the default uniform read distribution model provide similar level of accuracy with BitSeq performing slightly better than the other three methods.

Both BitSeq and Cufflinks use the same method for read distribution bias correction and provide improvement over the uniform model similar to improvements previously reported by \citet{Roberts2011}. 
We used version 0.9.3 of Cufflinks (as used by Roberts {\em et al.}) since we found that the most recent stable version of Cufflinks (version 1.3.0) leads to much worse performance for both uniform and bias-corrected models (see Supplementary results Section 2.2). 
The RSEM package uses its own method for bias correction based on the relative position of fragments, which in this case did not improve the expression estimation accuracy for the selected transcripts.
We were not able to compare the bias corrected results of MMSEQ \citep{Turro2011} due to an error in an external \texttt{R} package mseq used for the bias correction. 
However, the bias correction of mseq package itself was already compared against Cufflinks on the same dataset showing slightly worse accuracy and less improvement \citep{Roberts2011}.

In case of BitSeq, the major improvement of accuracy originates from using the effective length normalization.
To compare the results with qRT-PCR, the relative expression of fragments $\theta$ has to be converted into either relative expression of transcripts ($\theta^\ast$) or RPKM units.
Using the bias corrected effective length for this conversion leads to the higher correlation with qRT-PCR (Supplementary Table 1).
This means that using an expression measure adjusted by the effective length, such as RPKM, is more suitable than normalized read counts for DE analysis.

For more results comparing the transcript expression estimation accuracy and within gene relative expression accuracy, please refer to supplementary material Section 2.3.

\subsection{Differential expression analysis}

\begin{figure*}[!tpb]
   \centering
   \subfigcapskip-0.5ex
   \subfigcapmargin1ex   
   \subfigure[]{
   \includegraphics[width=0.23\textwidth]{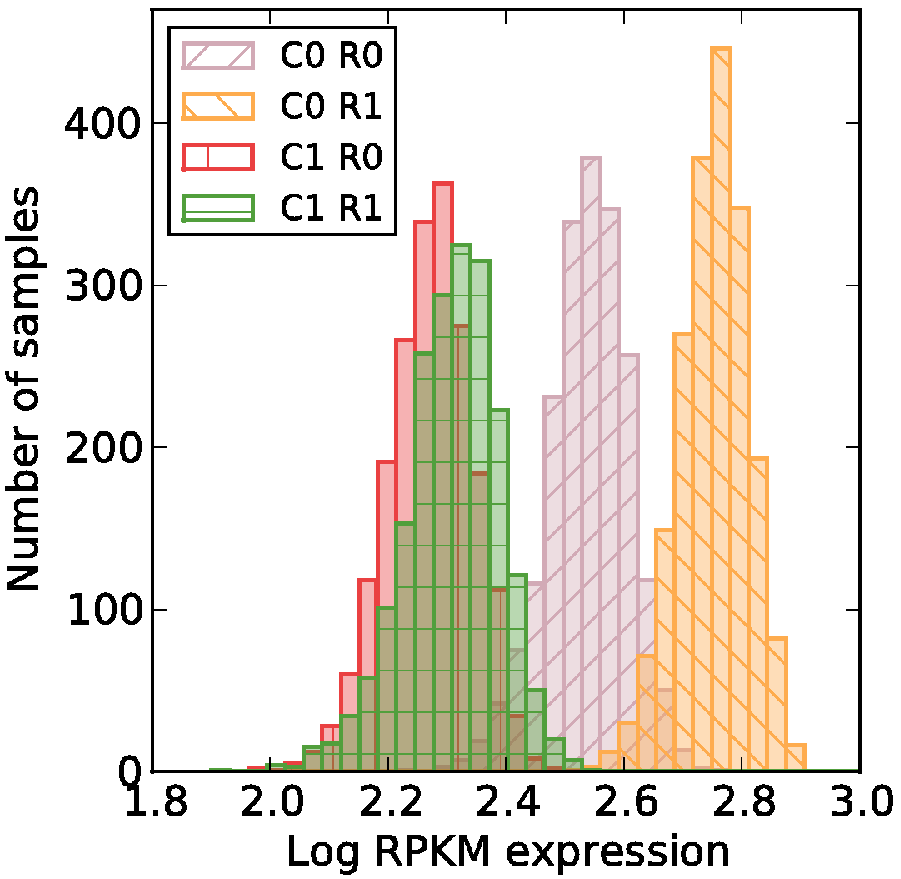}\label{fig:de1}}
   \subfigure[]{
   \includegraphics[width=0.23\textwidth]{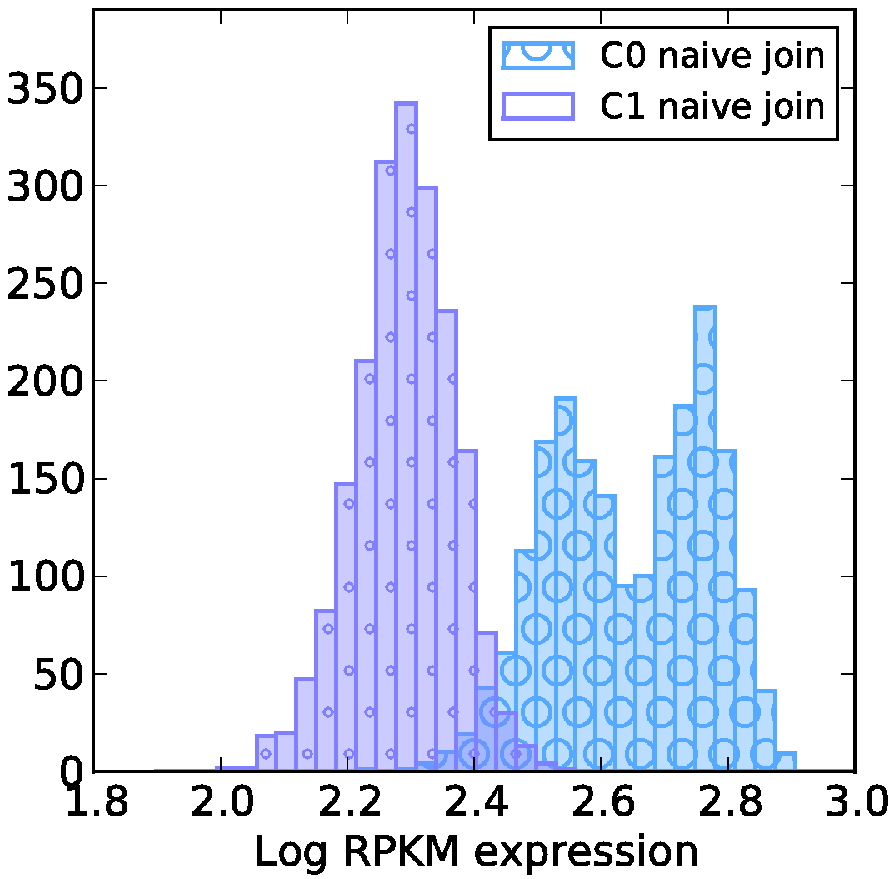}\label{fig:de2}}
   \subfigure[]{
   \includegraphics[width=0.23\textwidth]{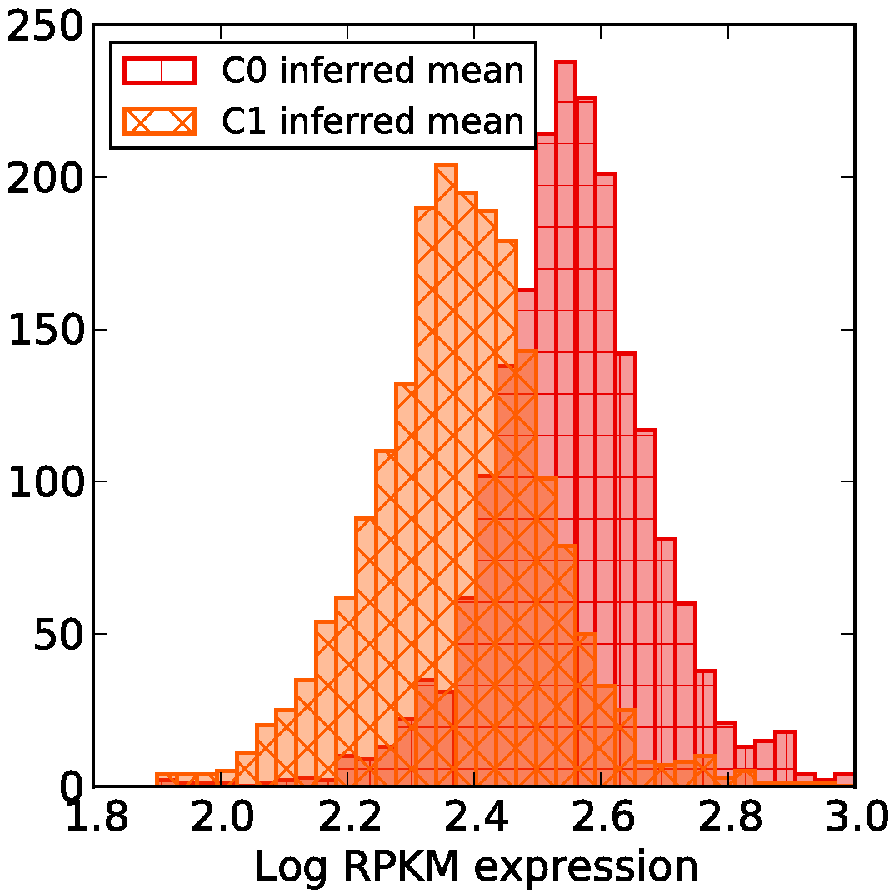}\label{fig:de3}}
   \subfigure[]{
   \includegraphics[width=0.23\textwidth]{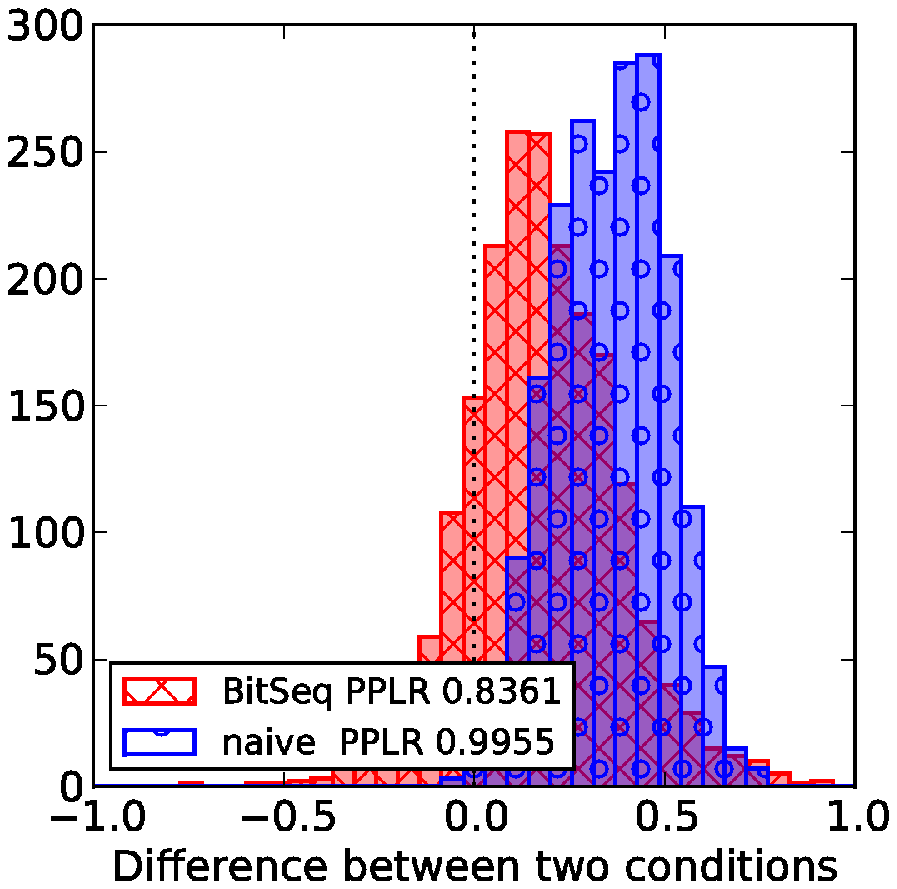}\label{fig:de4}}
\caption{ Comparison of BitSeq to naive approach for combining replicates within a condition for transcript uc001avk.2 of the Xu et al. dataset. 
(a) Initial posterior distributions of transcript expression levels for two conditions (labeled C0, C1), with two biological replicates each  (labeled R0, R1).  
(b) Mean expression level for each condition using the naive approach for combining replicates. The posterior distributions from replicates are joined into one dataset for each condition.
(c) Inferred posterior distribution of mean expression level for each condition using the probabilistic model in Figure~\ref{fig:modelDE}.
(d) Distribution of differences between conditions from both approaches show that the naive approach leads to overconfident conclusion.} \label{fig:de_dist}
\end{figure*}

We use the Xu et al. dataset to demonstrate the DE analysis process of BitSeq.
This dataset contains technical and biological replication for both studied conditions.  We observed significant difference between biological and technical variance of expression estimates  (Supplementary Figure~3).
Furthermore, the prominence of biological variance increases with transcript expression level.
We illustrate how BitSeq handles biological replicates to account for this variance in Figure \ref{fig:de_dist}, by showing the modelling process for one example transcript given only two biological replicates for each of two conditions.

Figure \ref{fig:de1} shows histograms of expression level samples produced in the first stage of our pipeline. 
BitSeq probabilistically infers condition mean expression levels using all replicates. 
For comparison, we used a naive way of combining two replicates by combining the posterior distributions of expression into a single distribution.
The resulting posterior distributions for both approaches are depicted in Figures \ref{fig:de2} and \ref{fig:de3}.

The probability of differential expression for each transcript is assessed by computing the difference in posterior expression distributions of the two conditions. 
Resulting distributions of differences for both approaches are portrayed in Figure \ref{fig:de4} with obvious difference in the level of confidence.
The naive approach reports high confidence of up-regulation in the second condition, with the probability of positive log ratio (PPLR) being $0.995$.
When biological variance is being considered by inferring the condition mean expression, the significance of differential expression is decreased to PPLR $0.836$.

\subsection{Assessing DE performance with simulated data}\label{sec:de_comparison}

\begin{figure}[!tpb]
   \centering
   \subfigcapskip-2.0ex
   \subfigcapmargin1ex
   \subfigure[]{
   \includegraphics[width=0.22\textwidth,trim=2mm 0mm 2mm 12mm,clip]{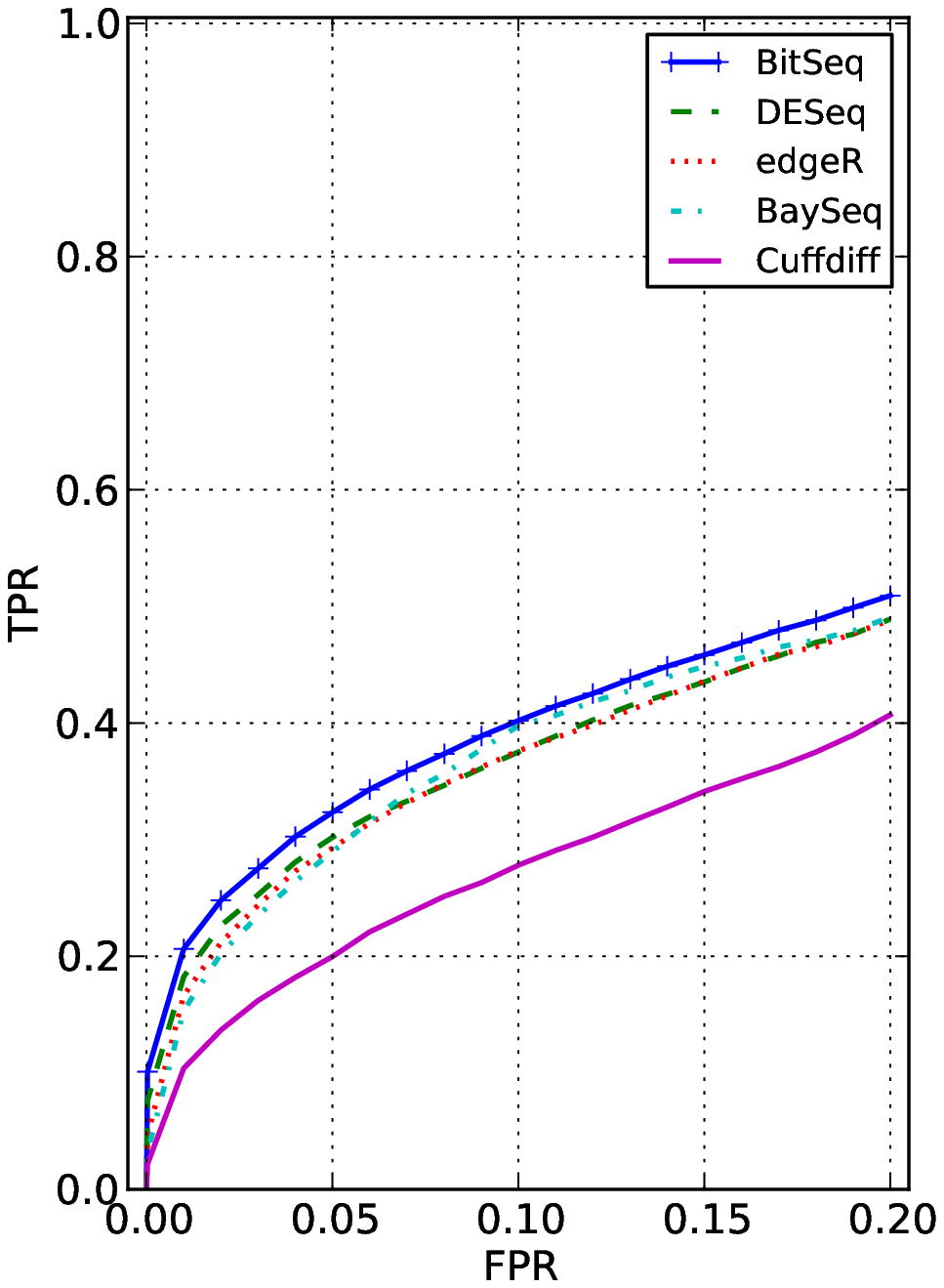}\label{fig:roc1-a}
   }
   \subfigure[]{
   \includegraphics[width=0.22\textwidth,trim=2mm 0mm 2mm 12mm,clip]{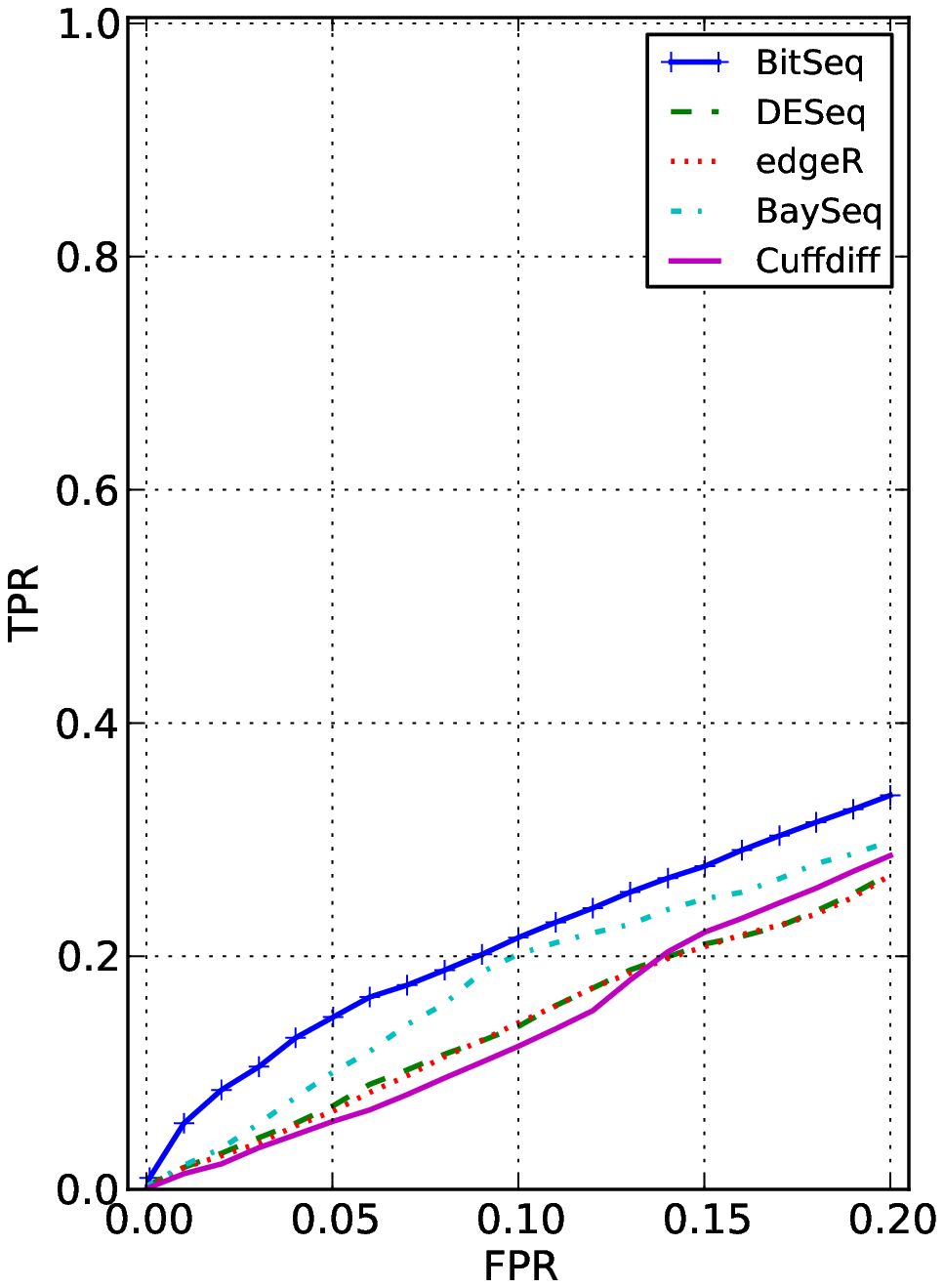}\label{fig:roc1-b}
   }
   \subfigure[]{
   \includegraphics[width=0.22\textwidth,trim=2mm 0mm 2mm 12mm,clip]{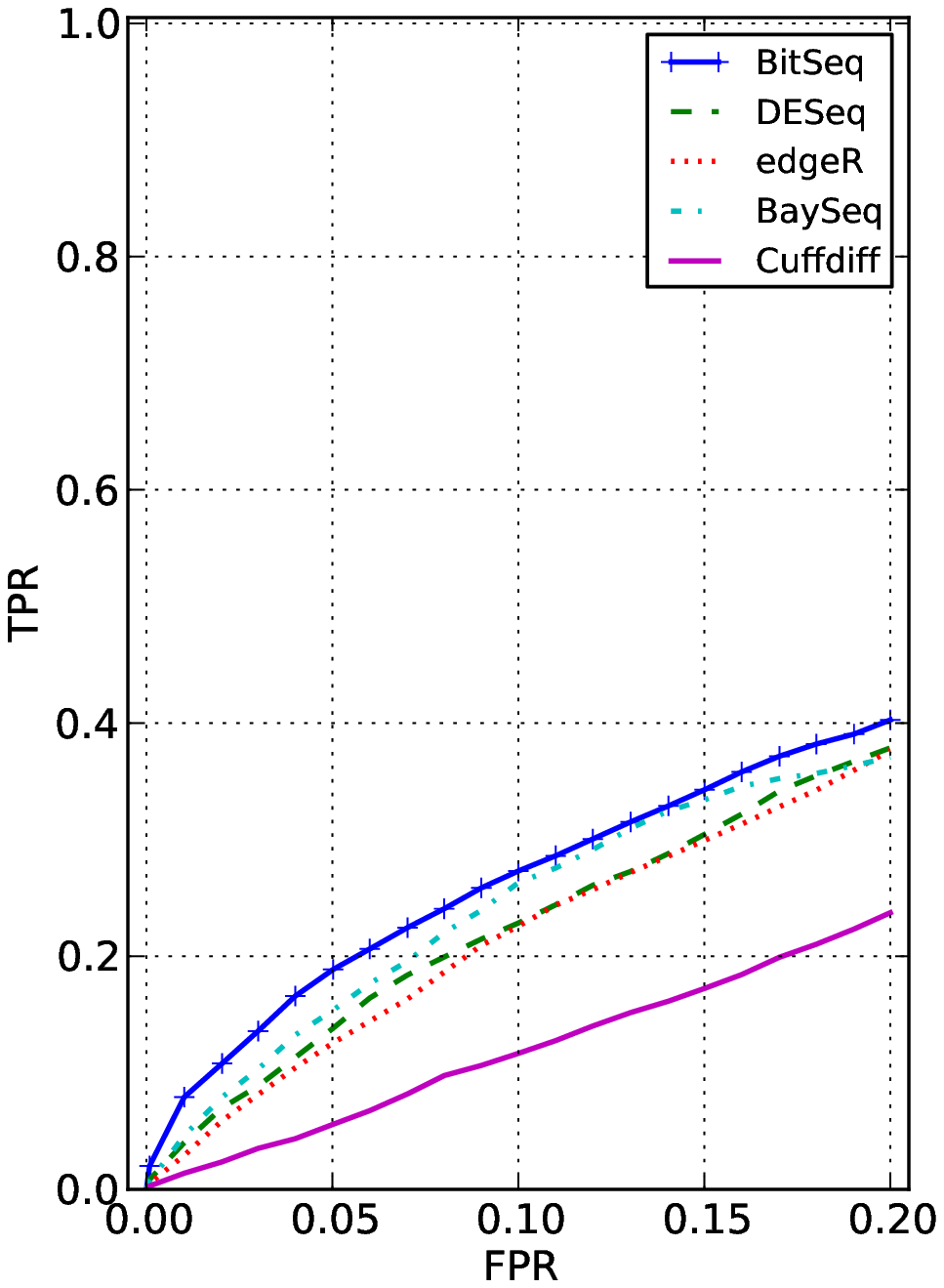}\label{fig:roc1-c}
   }
   \subfigure[]{
   \includegraphics[width=0.22\textwidth,trim=2mm 0mm 2mm 12mm,clip]{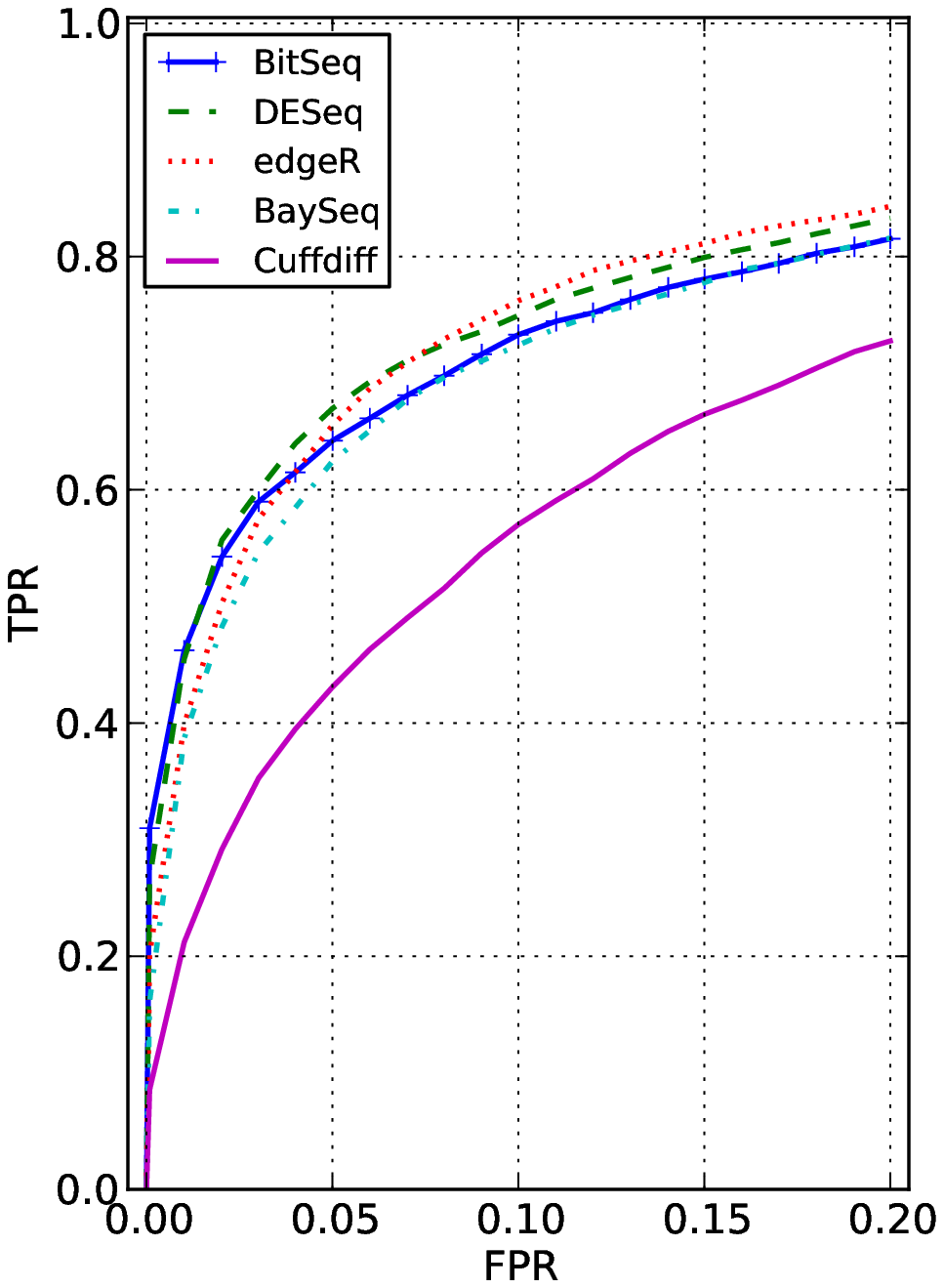}\label{fig:roc1-d}
   }
\caption{ROC evaluation of transcript level DE analysis using artificial dataset, comparing BitSeq with alternative approaches.
The curves are averaged over 5 runs with different set of transcripts being differentially expressed by fold change uniformly distributed in the interval $(1.5,3.5)$. 
We discarded transcripts without any reads initially generated as these provide no signal.
Panel (a) shows global average behaviour while in (b), (c) and (d)
transcripts were divided into 3 equally sized groups based on the logged mean generative read count: $[0,1.061)$, $[1.061,2.940)$, $[2.940,\infty)$, respectively.
}\label{fig:roc1}
\end{figure}

Using artificially simulated data with a predefined set of differentially expressed transcripts, we evaluated our approach and compared it with four other methods commonly used for differential expression analysis. DESeq \citep{Anders2010}, edgeR \citep{Robinson2010b}, baySeq \citep{Hardcastle2010} were designed to operate on the gene level and Cuffdiff \citep{Trapnell2010} on the transcript level.
Despite not being designed for this purpose, we consider the first three in this comparison as the use case is very similar and there are no other well known alternatives besides Cuffdiff that would use replicates for transcript level DE analysis.
All other methods beside Cuffdiff use BitSeq Stage 1 transcript expression estimates converted to counts.
Details regarding use of these methods are provided in the Supplementary material, Section 2.5. 
Figure \ref{fig:roc1} shows the overall results as well as split into three parts based on the expression of the transcripts.
The ROC curves were generated by averaging over 5 runs with different transcripts being differentially expressed and the figures are focused on the most significant DE calls with false positive rate below $0.2$. 

Overall (Figure \ref{fig:roc1-a}), BitSeq is the most accurate method,
followed first by baySeq, then edgeR and DESeq with Cuffdiff further
behind.  This trend is especially clear for lower expression levels
(Figure \ref{fig:roc1-b}, \ref{fig:roc1-c}).  The overall performance
here is fairly low because of high level of biological variance.  For
highest expressed transcripts (Figure \ref{fig:roc1-d}), DESeq and
edgeR show slightly higher true positive rate than BitSeq and baySeq,
especially at larger false positive rates.
Further details and more results from the DE analysis comparison can be found in the supplementary material Section 2.5.

%% file: conclusion.tex
\section{Conclusion}

We have presented methods for transcript expression level analysis and differential expression analysis that aim to model the uncertainty present in RNA-seq datasets. 
We used a Bayesian approach to provide a probabilistic model of transcriptome sequencing and to sample from the posterior distribution of the transcript expression levels.
The model incorporates read and alignment quality, adjusts for non-uniform read distributions and accounts for experiment-specific fragment length distribution in case of paired-end reads.
The accuracy of inferred expression is comparable and in some cases outperforms other competing methods.
Nevertheless, the major benefit of using BitSeq for transcript expression inference is the availability of full posterior distributions useful for further analysis.

The inferred distributions of transcript expression levels can be further analysed by the second stage of BitSeq for DE analysis.
Given biological replicates, BitSeq accounts for the intrinsic noise and variation and produces more reliable estimates of expression levels within each condition, thus providing fewer false differential expression calls.
We want to highlight that in order to make most accurate differential expression assessment, experimental design must include biological replication.
BitSeq is capable of combining information from multiple biological and technical replicas and comparing multiple conditions.
Further studies including multiple replicates are necessary to investigate the effects of library preparation and biological variance.


%% file: supplement_body.tex
\newcommand{\B}[1]{\begin{#1}}
\newcommand{\E}[1]{\end{#1}}
\newcommand{\TP}{\theta'}
\newcommand{\TAA}[1]{\theta^{act#1}}
\newcommand{\ZZ}[1]{Z_{n#1}}
\newcommand{\ZZA}{\mathscr{Z}^{act}}
\newcommand{\Leff}{l_m^{\textrm{\tiny eff}}}
\newcommand{\MM}[1]{\mathscr{#1}}
\newcommand{\RR}{\mathscr{R}}
\newcommand{\BS}[1]{\boldsymbol{#1}}

\newcommand{\BI}{\begin{itemize}%
   \setlength{\itemsep}{1pt}%
   \setlength{\parskip}{0pt}%
   \setlength{\parsep}{0pt}}
\newcommand{\EI}{\end{itemize}}
\newcommand{\BE}{\begin{enumerate}}
\newcommand{\EE}{\end{enumerate}}
\newcommand{\KOMENTAR}[1]{}

\newcommand{\LAC}{{\lambda_m^{(c)}}}
\newcommand{\LACR}{{\lambda_m^{(cr)}}}
\newcommand{\LAZ}{{\lambda_0}}
\newcommand{\MUC}{{\mu_m^{(c)}}}
\newcommand{\MUZ}{{\mu_m^{(0)}}}
\newcommand{\YEX}[1]{{y_m^{(#1)}}}

\hyphenation{tran-script-ome}

\section{Methods}

\subsection{Alignment probabilities}

We present the alignment probability computation for the case of
paired end reads.  For single reads, the terms related to fragment or
insert length distribution and the other paired read disappear.

For a given transcript $I_n =m; m\in\{1,\dots, M\}$, the probability of observing a pair of reads $(r_n^{(1)}, r_n^{(2)})$ is determined by the probability of the read being sequenced from a specific strand $s$ at a specific position $p$ with a specific insert length $l$ and the probability of reporting the reads after sequencing the sequences $(seq_{mlps}^{(1)}, seq_{mlps}^{(2)})$, 
\begin{equation}\label{eq:alignmentProbability_appendix}
P(r_n^{(1)},r_n^{(2)}|I_n=m) = \\
P(l|m)P(p|l,m)P(s|m)P(r_n^{(1)}|seq_{mlps})P(r_n^{(2)}|seq_{mlps}) \ .
\end{equation}
Unless a strand specific sequencing protocol is used, the probability
of observing a read from either strand is the same, $P(s|m)=1/2$, and
can be ignored.
The fragment length distribution $P(l|m)$ is assumed to be log-normal
with its parameters given by the user or estimated from read pairs
with only a single transcript alignment.

The probability of sequencing a given position is in general
\begin{equation}
  \label{eq:positionProb}
  P(p|I_n = m, l) = \frac{b_m(p)}{ \sum_{p=1}^{l_m-l_r+1} b_m(p)}.
\end{equation}
where $b_m(p)$ denotes bias for a particular position $p$ on
transcript $m$.
For a constant $b_m(p)$ corresponding to a uniform read distribution,
this reduces to 
$P(p|m)=1/(l_m-l_r+1)$ which only depends on the
lengths of the transcript $l_m$ and the read $l_r$.

We calculate the probability of observing a sequence based on the read's quality base scores and mismatches.

The \textit{Phred} score can be converted into probability of
base-calling error $p_{\text{err},i}$. The final sequence probability
is now obtained as
\begin{equation}
  \label{eq:seqProb}
  P(r_n^{(j)}|seq_{mps}) = \prod_{i \in \mathrm{matches}} (1-p_{\text{err},i})
  \prod_{i \in \mathrm{mismatches}} p_{\text{err},i},
\end{equation}
where the probability of error for a given base $i$ is based on the
Phred score $p_{\text{err},i} = 10^{-\text{Phred}_i/10}$.

\subsubsection{Bias estimation}

Our model can easily incorporate a correction for position and
sequence specific biases.  One example of such a model is presented by
\citet{Roberts2011} for correcting the fragmentation bias.
Under this model, we have
\begin{equation}
  \label{eq:sequenceBias}
  b_m(p) = b^{s,5}_m(e_5) b^{s,3}_m(e_3) b^{p,5}_m(e_5) b^{p,3}_m(e_3),
\end{equation}
where $b^{s,5}_m(e_5)$ and $b^{s,3}_m(e_3)$ are the sequence specific
biases for 5' and 3' ends of the fragment, respectively, and
$b^{p,5}_m(e_5)$ and $b^{p,3}_m(e_3)$ are the corresponding positional
biases.

We use separate variable length markov models to capture the bias for
each end.  The structure of this model is the same as that
of~\citet{Roberts2011}, presented in Figure 2 of the supplementary
methods.  For the sequence bias these are
\begin{equation}
  \label{eq:sequenceBias2}
  b^{s,5}_m(e_5) = \prod_{n=1}^{21}\frac{ \psi^{5,R}_{n,\pi_n}}{\psi^{5,U}_{n,\pi_n}},
\end{equation}
which are based on 21 probabilities $\psi^5_{n,\pi_n}$ from 8 bases
before and 12 bases after the read starting position.  Here $\psi^{5,R}$
refers to the biased and $\phi^{5,U}$ to a uniform model, $n$ is a
node or a position, $\pi_n$ are the parents of node $n$ and
$\psi^5_{n,\pi_n}$ is the probability of base $X$ at node (or
position) $n$ given the bases observed on parent nodes $\pi_n$.
The model has 744 parameters in all, with each node having 0, 1 or 2
parents as in the model of~\citet{Roberts2011}.
The parameters are estimated from empirical frequencies using reads
with a single alignment.
For a read $r$ aligning to transcript $m$ we increase appropriate
probabilities $\psi^{5,R}$ by $1/\theta_m$, where $\theta_m$ is an
initial coarse expression estimate obtained by running BitSeq with uniform read distribution model beforehand.
In the contrasting uniform model for all $K=l_m-l_r+1$ possible
positions of read of length $l_r$, the appropriate
probabilities $\psi^{5,U}$ are increased by $\frac{1}{\theta_m K}$.
The model $b^{s,3}_m(e_3)$ is similar.


In addition to the sequence-specific bias, there is a model for
positional bias within the transcript.  This is
\begin{equation}
  \label{eq:posBias}
  b^{p,5}_m(e_5) = \frac{\omega_{l_m,e_5/l_m}^{R}}{\omega_{l_m,e_5/l_m}^{U}},
\end{equation}
where $\omega_{l,p}$ is the probability for starting position within
transcript of length $l$ on position $p$.  The probabilities are
modelled within 5 transcript length bins and 20 bins of relative
position.  The probabilities are again estimated from empirical
frequencies of reads with single alignments taking into account
expression $\theta$.

\subsection{Effective length computation}

For the purpose of reporting normalized measure such as RPKM, $\boldsymbol{\theta}$, the relative expression of fragments, has to be normalized by the amount of reads or fragments that can be produced by a unit of transcript.
When assuming uniform read distribution of single-end reads, this would be $l_m-l_r$ as the number of starting positions for a read of length $l_r$.
For pair-end reads, the effective length of a transcript has to account for fragment length distribution as well,
\begin{equation}
   \label{eq:effLength}
   l^{(eff)}_m = \sum_{l_f=1}^{l_m} p(l_f|m) * (l_m-l_f).
\end{equation}

With the use of read distribution with bias correction, we learn more about the distribution of fragments and thus can use this information when computing the effective length.
In this case, the effective length takes into account bias weight for every position of the transcript,
\begin{equation}
   \label{eq:effLengthBias}
   l^{(eff+bias)}_m = \sum_{l_f=1}^{l_m} p(l_f|m) \sum_{p=1}^{l_m-l_f}  b_m(p)
\end{equation}
As we show later in Section \ref{sec:bias} of this Supplementary material, using the bias corrected effective length can substantially improve the accuracy of our method.

\subsection{Gibbs sampling in expression estimation (Stage 1)}

We apply a collapsed Gibbs sampler for Stage 1 estimation by
marginalising out the expression level and noise level parameters
$\boldsymbol{\theta}$ and $\TA$ and iteratively resampling the isoform
assignments $I_n$ of each read given the assignments of other reads
$I^{(-n)}$.  The full update rules for the sampler are
\begin{align}\label{eq:collapsed}
P(I_n|I^{(-n)}, R) & = \textrm{Cat}(I_n|\boldsymbol{\phi_n^{\ast}}) , \\
\phi_{n0}^{\ast} & = P(r_n|\textrm{noise}) (\beta^{act} + C_0^{(-n)}) /Z^{(\phi^\ast)}_n ,\notag\\
m\neq 0; \phi_{nm}^{\ast} & = P(r_n|I_n) (\alpha^{act} + C_+^{(-n)})
  {\textstyle \frac{(\alpha^{dir} + C_m^{(-n)})}{(M\alpha^{dir} + C_+^{(-n)})}}  / Z_n^{(\phi^\ast)} ,\notag\\
   C_m^{(-n)} & = {\textstyle \sum_{i\neq n}} \delta(I_i=m) ,\notag\\
   C_+^{(-n)} & = {\textstyle \sum_{i\neq n} \delta(I_i> 0)} \ ,\notag
\end{align}
with $Z_n^{(\phi^\ast)}$ being a constant normalising $\boldsymbol{\phi_n}^{\ast}$
to sum up to 1, and $\alpha^{dir}=1 ,\alpha^{act}=2,\beta^{act}=2$.

As an alternative, it is also possible to use a regular Gibbs sampler
alternating between sampling $I_n$ and $\boldsymbol{\theta}$.  The
corresponding update rules are
\begin{align}
P(I_n|\boldsymbol{\theta},\TA,R) & = \textrm{Cat}(I_n|\boldsymbol{\phi_n}) , \\
\phi_{n0} & = P(r_n|\textrm{noise})(1-\TA) / Z_n^{(\phi)}, \notag\\
m\neq 0; \phi_{nm} & = P(r_n|I_n)\theta_m\TA / Z_n^{(\phi)}, \notag\\
P(\boldsymbol{\theta}|\boldsymbol{I},\TA,R) & = \textrm{Dir}(\boldsymbol{\theta}| (\alpha^{dir} + C_1,\dots ,\alpha^{dir} + C_M)), \\
P(\TA|\boldsymbol{I},\boldsymbol{\theta},R) & = \textrm{Beta}(\TA| \alpha^{act} + N - C_0, \beta^{act} + C_0), \\
C_m & = {\textstyle \sum_{n=1}^{N} } \delta(I_n=m) . \notag
\end{align}
This approach is usually less efficient in practice, though.

\subsection{Differential Expression model (stage 2)}

The Differential Expression (DE) model is shown in Figure~3 of the main paper. 
We consider data from conditions $c=1\dots C$ with number of replicates for 
each condition denoted $R_1,\dots,R_C$. We fit the model to each transcript $m$
 independently using 
``pseudo-data" $y_m^{(cr)} = \log \theta_m^{(cr)}$ which is created from MCMC
samples from Stage 1. One sample of $\theta_m^{(cr)}$ is drawn for each $(r,c)$
combination to create a pseudo-data vector $\bm y_m$ of length $\sum_{c=1}^C {R_c}$.
Inference is carried out independently for each pseudo-data vector and 
the results are then combined as described in the main text. This allows the 
technical error from Stage 1 to be propagated through the model. Since the model
is conjugate then the inference for each pseudo-data vector is exactly tractable
and no further MCMC is required to sample the condition means.

\subsubsection{Parameter estimation for each transcript}

The condition means are denoted $\boldsymbol{\mu_m} = (\mu_m^{(1)},\dots,\mu_m^{(C)})$
and we are interested in inferring the posterior distribution over the means given one 
pseudo-data vector $\bm y_m$. The model is defined as,
\begin{align*}
y_m^{(cr)} \sim &~ \textrm{Norm} (\mu_m^{(c)},1/\lambda_m^{(c)}) \\
\mu_m^{(c)} \sim &~ \textrm{Norm} (\mu_m^{(0)}, 1/(\lambda_0 \lambda_m^{(c)} )) \\
\lambda_m^{(c)} \sim &~ \textrm{Gamma} (\alpha_G, \beta_G)
\end{align*}
with hyper-parameters $\lambda_0,\alpha_G,\beta_G$ which are estimated from 
groups of transcripts with similar mean expression across conditions. The
 hyper-parameter $\mu_m^{(0)}$ is fixed at the empirical mean transcript expression
across conditions.


\begin{align*}
p(\BS{\mu}_m, \BS{\lambda}_m|\BS{y}_m) &
\propto p(\BS{y}_m| \BS{\mu}_m, \BS{\lambda}_m )p(\BS{\mu}_m)p(\BS{y}_m) \\
& \propto \prod_{c=1}^C p(\MUC)p(\LAC) \prod_{r=1}^{R_c}  p(\YEX{cr}| \MUC, \LAC ) \\
%
%
%
& \propto \prod_{c=1}^C  \textrm{Gamma}(\LAC| a_c, b_c ) 
\textrm{Norm}\Brack{\mu_c \left| \frac{\LAZ\MUZ+Syc}{\LAZ+R_c}, \frac{1}{\LAC(\LAZ+R_c)} \right.} \\
a_c & = \alpha_G + \frac{R_c}{2} \\
b_c & = \beta_G +\frac{1}{2}\Brack{ \LAZ\MUZ^2 + S^2yc -\frac{(\LAZ\MUZ+Syc)^2}{\LAZ+R_c}}
\end{align*}
where $Syc$ denotes $\sum_{r=1}^{R_c} \YEX{cr}$ and $S^2yc$ denotes $\sum_{r=1}^{R_c} \YEX{cr}^2$.

\subsubsection{Hyper-parameter estimation across transcript groups}

For hyper-parameter estimation we consider a set of transcripts $m=1\dots M'$ in a group $g$ of transcripts with similar expression. The hyperparameter $\mu_0$ is now set to the mean expression per group of transcripts  and $\lambda_0$ is set to 2.0. We have
pseudo-data samples $\YEX{cr}$ for each transcript and we are interested in hyperparameters $\alpha$ and $\beta$, where $\beta$ is the rate of Gamma distribution. The model is defined as,
\begin{align*}
\YEX{cr} \sim & \textrm{Norm}(\MUC, 1/\LAC ) \\
\MUC \sim & \textrm{Norm}(\MUZ, 1/(\LAC\LAZ)) \\
\LAC \sim & \textrm{Gamma}(\alpha, \beta) \\
P(\alpha,\beta) \sim & \textrm{Uniform}(0,\infty)
\end{align*}
The hyper-parameter posterior distribution is given by,
   

\begin{align*}
P(\alpha, \beta | \BS{y} ) & \propto P(\alpha,\beta)P(\BS{y}|\alpha,\beta) \\
& \propto \prod_{m=1}^{M'} \prod_{c=1}^{C} P(\BS{y}_m^{c}|\alpha,\beta) \\
& \propto \prod_{m=1}^{M'} \prod_{c=1}^{C} \int d\LAC p(\LAC|\alpha,\beta)
\int d\MUC P(\MUC|\LAC) \prod_{r=1}^{R_c} P(\YEX{cr}|\LAC, \MUC)\\
%
& \propto  \prod_{m=1}^{M'} \prod_{c=1}^{C} \frac{\beta^\alpha}{\Gamma(\alpha)}
\frac{\Gamma(\alpha+R_c)}{ \Brack{\beta+\frac{1}{2}\Brack{\LAZ\MUZ^2 + S^2yc - \frac{(\LAZ\MUZ+Syc)^2}{\LAZ+R_c}}}^{\alpha+R_c}} \ . \\
\end{align*}
\noindent This distribution is not in a standard form and we use Metropolis-Hastings Random walk MCMC to sample $\alpha$ and $\beta$. We then use lowess smoothing across groups to estimate the mean hyper-parameter for each transcript according to its empirical mean expression level across conditions.


\section{Results}

\subsection{Transcript expression inference}

\begin{figure}[!tpb]
   \centering
   \subfigcapskip-0.5ex
   \subfigcapmargin1ex
   \subfigure[Anti-correlation of transcripts.]{
   \includegraphics[width=0.3\textwidth]{corr-1678b.eps}
   }%
   \subfigure[No observable correlation.]{
   \includegraphics[width=0.3\textwidth]{corr-1678a.eps}
   }
   \subfigure[Anti-correlation of transcripts..]{
   \includegraphics[width=0.3\textwidth]{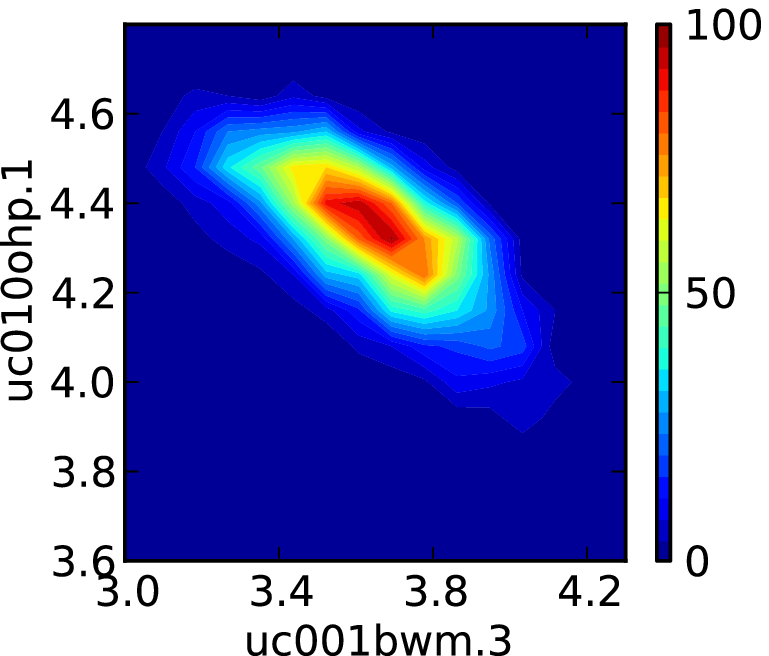}\label{fig:corr-c}
   }
   \subfigure[Transcript sequence profile.] {
   \includegraphics[width=0.98\textwidth]{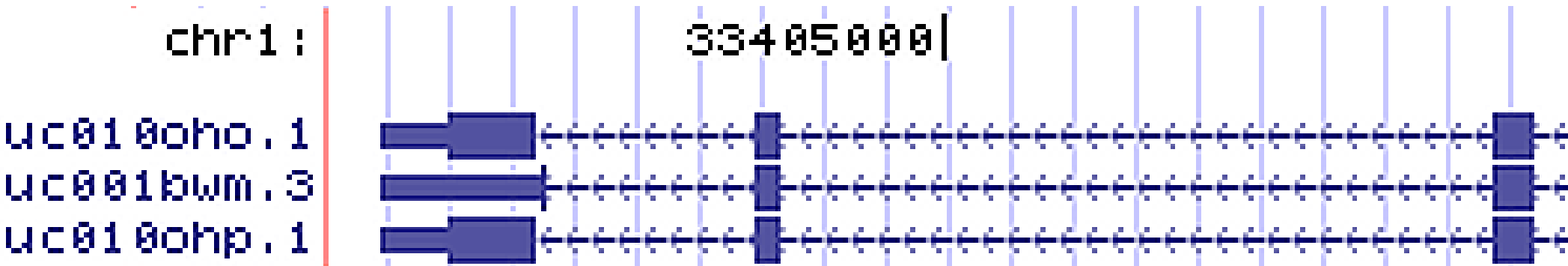}\label{fig:prof}
   }
   \caption{In plots (a), (b) and (c) we show the posterior transcript expression density for pairs of transcripts from the same gene. This is a density map constructed using the MCMC expression samples for these three transcripts. In (d) we show transcript sequence profile obtained from the UCSC genome browser. 
   The sequencing data is from miRNA-155 study published by \citet{Xu2010a}.
}\label{fig:cor_hist2}
\end{figure}

In the main text (Figure 4) we illustrate the correlation present in the expression posterior distribution for transcripts that share a large proportion of transcribed sequence.
Here we provide all three pairwise plots for the transcripts  uc010oho.1, uc010ohp.1 and uc001bwm.3, which are the only transcripts of gene Q6ZMZ0 in the UCSC known Gene annotation.
The expression samples of uc001bwm.3 and uc010ohp.1 (Figure \ref{fig:corr-c}) are also negatively correlated. 
This means that the model is not able to decide from which transcript some of the reads originated and the posterior distribution captures all viable assignments.
The transcript sequence profile in Figure \ref{fig:prof} clearly demonstrates the similarity of the transcripts that causes higher uncertainty when inferring the transcript expression levels.

\subsection{Read distribution bias correction}\label{sec:bias}

\begin{table}[!tpb]
   \centering
\begin{tabular}{l|c||c|c|c|c}
Method ver. & Read distribution & Average & Rep. 1 & Rep. 2 & Rep. 3 \\
\hline
\hline
BitSeq 0.4 &uniform $~^\ast$ & 0.7585 & 0.7575 & 0.7580 & 0.7594 \\
BitSeq 0.4 & uniform $~^\dagger$ & 0.7677 & 0.7672 & 0.7669 & 0.7675 \\
BitSeq 0.4 & bias corrected $~^\ast$ & 0.7565  &0.7554 & 0.7561 & 0.7573  \\
BitSeq 0.4 & bias corrected $~^\dagger$  & 0.7652 & 0.76495 & 0.7647 & 0.7652 \\
BitSeq 0.4 & bias corrected $~^\ddagger$ & \textbf{0.8011} & 0.8018 & 0.7959 & 0.8041 \\
\hline
Cufflinks 0.9.3 &uniform & 0.7503 & 0.7470 & 0.7513 & 0.7519 \\
Cufflinks 0.9.3 &bias corrected & \textbf{0.8056} & 0.8018 & 0.8050 & 0.8083 \\
Cufflinks 1.3.0 &uniform & 0.5331 & 0.5130 & 0.5336 & 0.5477 \\
Cufflinks 1.3.0 &bias corrected & 0.6842 & 0.6858 & 0.6917 & 0.6446 \\
\hline
RSEM 1.1.14 &uniform & 0.7632 & 0.7623 & 0.7628 & 0.7640  \\
RSEM 1.1.14 &bias corrected & \textbf{0.7633} & 0.7623 & 0.7628 & 0.76409  \\
\hline
MMSEQ 0.9.18 & uniform & \textbf{0.7614} & 0.76099 & 0.7606 & 0.7620 \\
\end{tabular}

\caption{Evaluation of transcript expression inference algorithms using the SRA012427 RNA-seq data and TaqMan qRT-PCR expression measures for 893 matching transcripts. 
Reported values are Pearson $R^2$ correlation coefficient of the 893 transcripts' expression estimates and qRT-PCR results, best correlation of a method using averaged expression is highlighted. 
For each method we present values for average expression taken from three replicates as well as for each technical replicate separately.
BitSeq was used with three different versions of expression length normalisation: $\ast$ -- using actual transcript length, $\dagger$ -- using effective length accounting for fragment length distribution, $\ddagger$ -- using effective length accounting for fragment length and read distribution bias.
}\label{tab:exp2}
\end{table}

We compared four different methods for expression estimation which include bias correction options for non-uniform read distribution. 
The extended results are presented in Table \ref{tab:exp2}, where we report the $R^2$ correlation of 893 transcript expression estimates with the TaqMan qRT-PCR results.
We used every method to analyse each of the three technical replicates separately and then used the average expression level for the comparison. 
As was already stated in the main paper, the newest stable version of Cufflinks does provide the lowest correlation. 
We resorted to using the version 0.9.3 which was used in the paper presenting the bias correction method adopted by BitSeq \citep{Roberts2011}.

For BitSeq, the major benefit of the bias correction algorithm comes from the effective 
transcript length normalisation. 
Relative expression of fragments used by BitSeq can be converted into relative expression of transcripts or into RPKM measure by adjusting the expression by effective length (see Supplementary Section 1).
In Table \ref{tab:exp2} we compare three different approaches for length normalisation.
In the first approach ($\ast$), the expression is adjusted by the length of a transcript. 
The second approach ($\dagger$) uses effective length taking into account the paired-end read fragment length distribution and the number of all positions from which a fragment could originate.
The last approach ($\ddagger$), which provides best results on this dataset, uses effective length computed using the fragment length distribution as well as read distribution bias weights (see Equation \ref{eq:effLengthBias}).
More careful investigation of this process is required, however it is limited by small number of RNA-seq datasets with known underlying expression, especially when using paired-end reads.

\subsection{Assessing transcript expression inference using simulated data}

\begin{figure}[!tpb]
   \centering
   \subfigcapskip-2.0ex
   \subfigcapmargin1ex
   \subfigure[]{
   \includegraphics[width=0.43\textwidth]{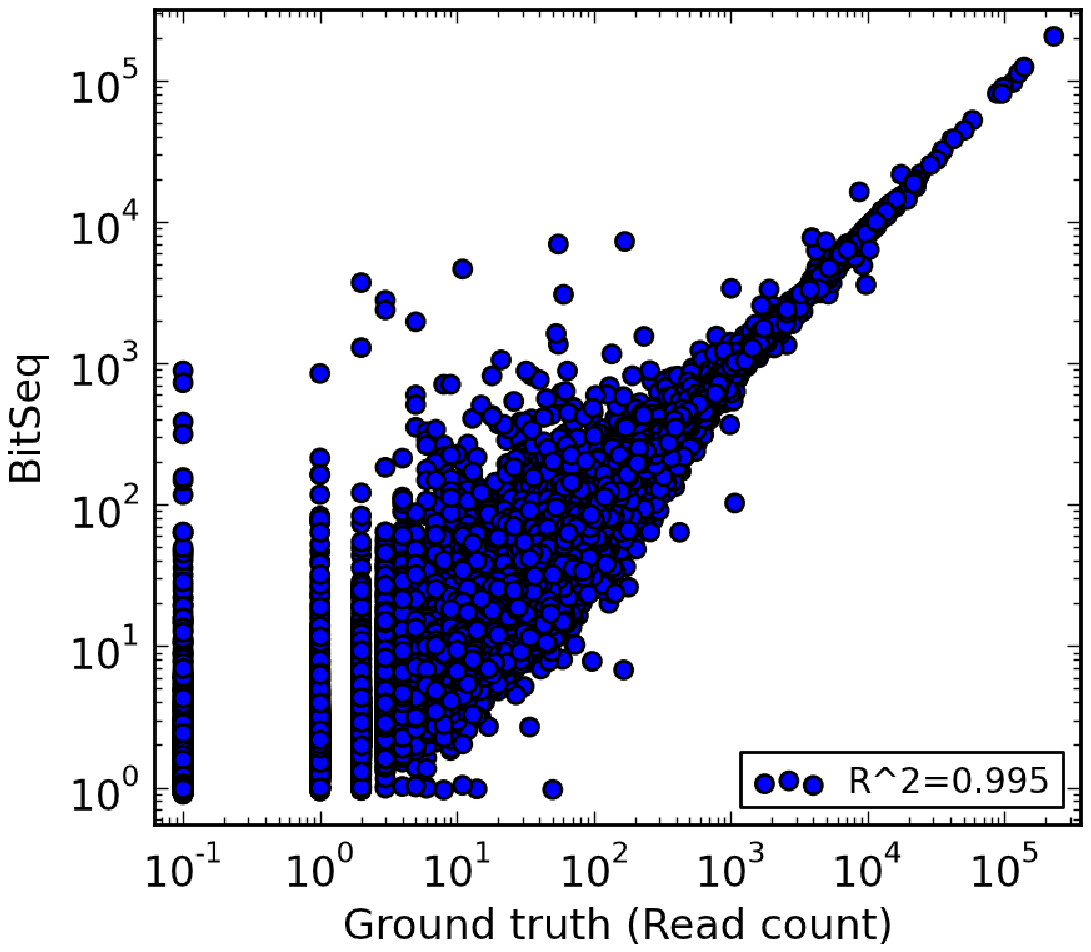}\label{fig:sim1-a}
   }
   \subfigure[]{
   \includegraphics[width=0.430\textwidth]{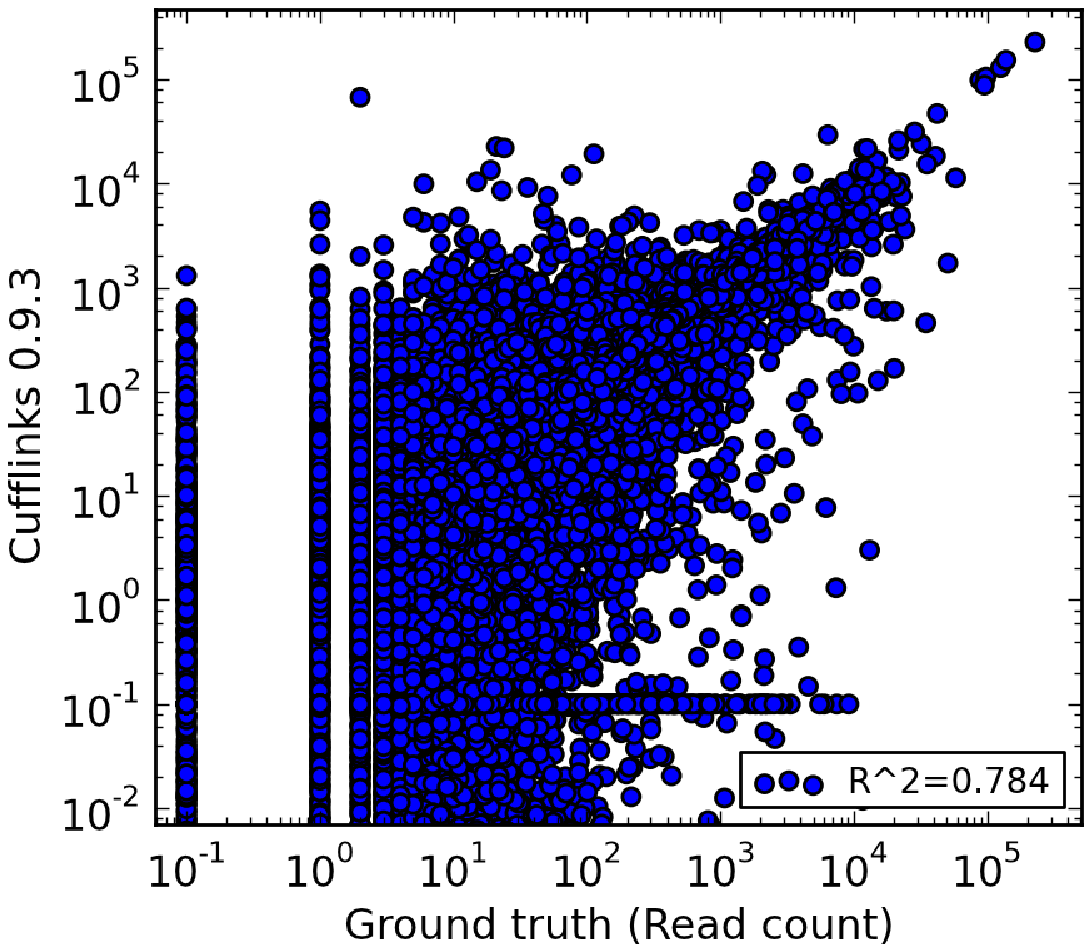}\label{fig:sim1-b}
   }
   \subfigure[]{
   \includegraphics[width=0.43\textwidth]{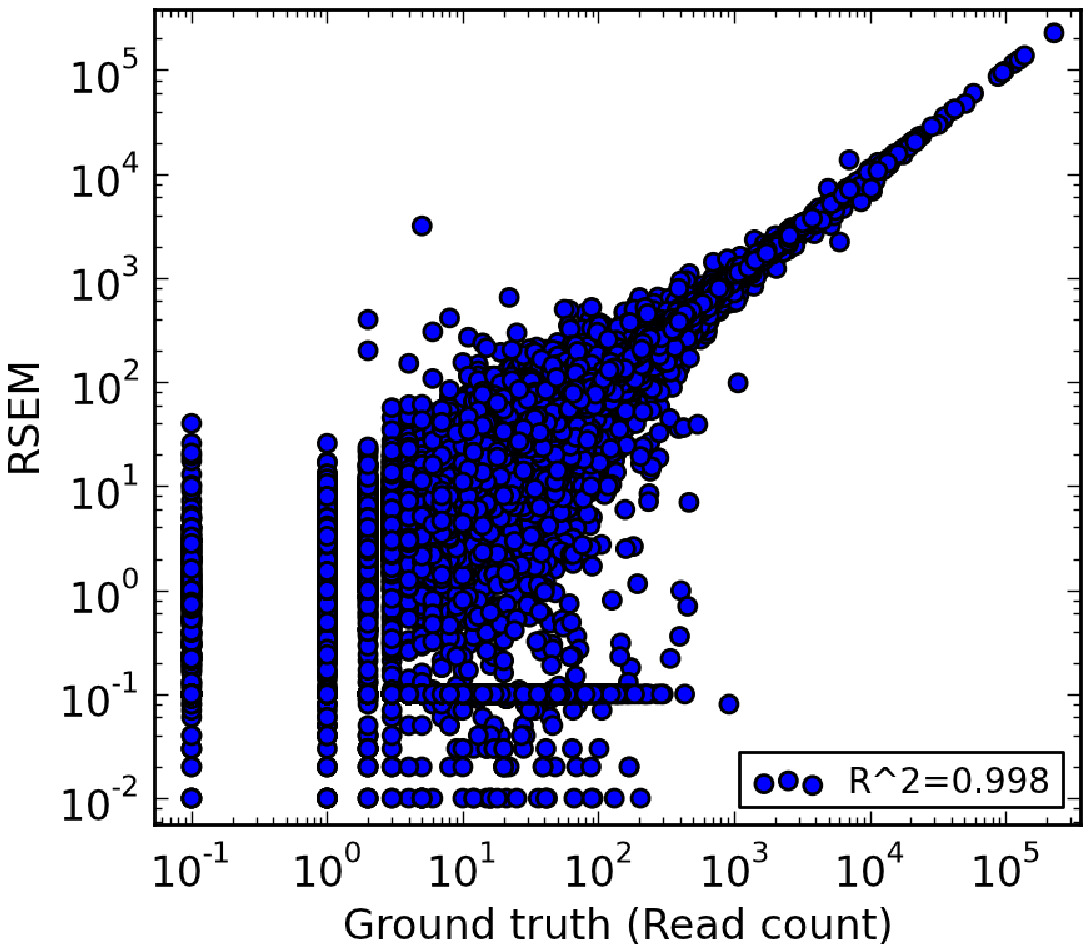}\label{fig:sim1-c}
   }
   \subfigure[]{
   \includegraphics[width=0.43\textwidth]{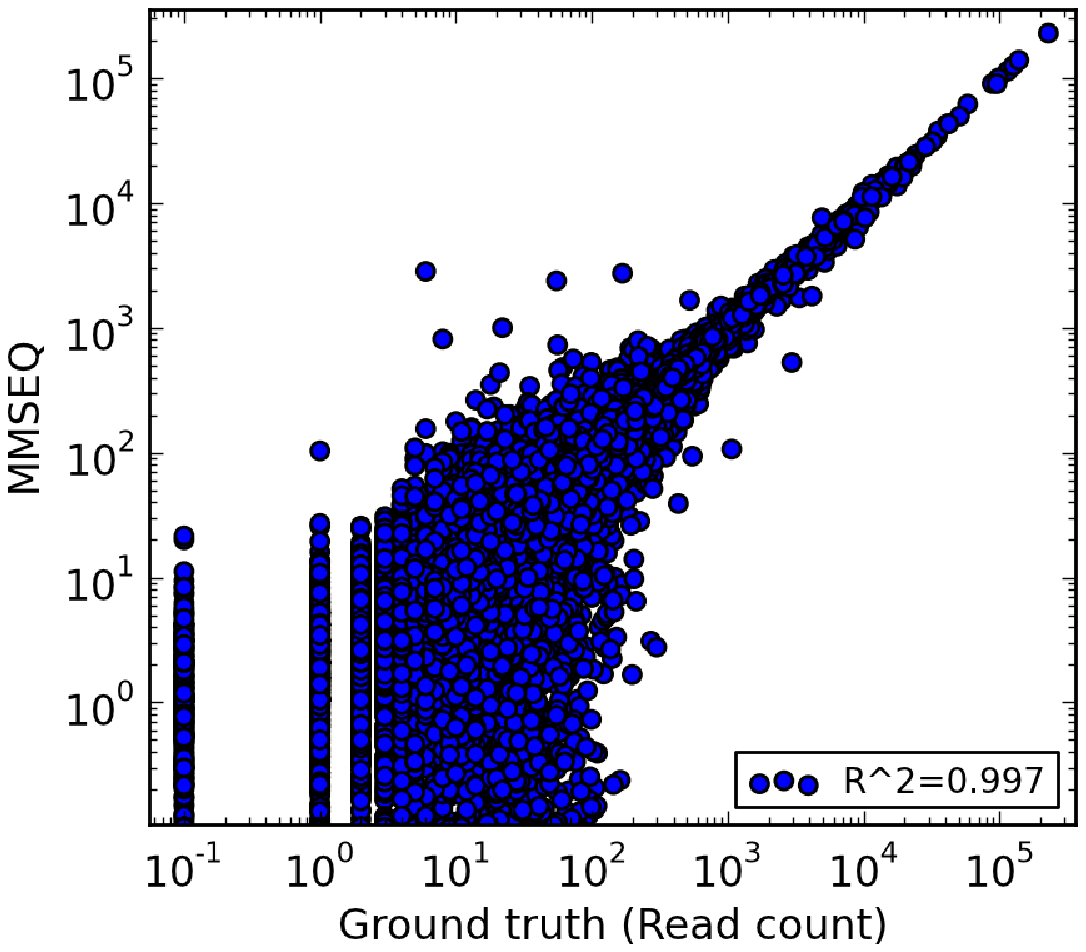}\label{fig:sim1-d}
   }
\caption{Comparison of expression estimates using 10M simulated pair-end reads with known expression.
The expression estimates were converted into estimated read counts for each transcript and compared against ground truth using Log-Log plot.
We calculated Pearson $R^2$ correlation coefficient for transcripts with at least one generated paired-end read.The figures show (a) BitSeq, (b) Cufflinks v0.9.3, (c) RSEM, and (d) MMSEQ.
}\label{fig:sim}
\end{figure}

We used simulated dataset of 10M paired-end reads to examine the expression estimation accuracy of BitSeq and compared it against other three popular methods. 
The reads were generated based on expression estimates from the Xu et al. dataset with a fragment size distribution $l_f\sim\text{LogNorm}(5.32, 0.12)$, inferred from the SRA012427 dataset. 
We used the UCSC NCBI37/hg19 knownGene annotation transcripts to generate the read fragments.
First we compared the overall expression accuracy against the generative read count values (Figure \ref{fig:sim}).
We used read count in order to facilitate the second part of our comparison, the assessment of the within gene relative expression estimation (Table \ref{tab:relative}), for which the use RPKM would not be feasible.

For comparison of overall expression accuracy, we report the Pearson $R^2$ correlation coefficient with the ground truth. 
The coefficient was calculated for transcripts with at least one read generated (46841 transcripts). In this comparison RSEM ($R^2=0.998$) has the highest correlation with MMSeq ($R^2=0.997$) and BitSeq ($R^2=0.995$) being closely behind. 
Unfortunately we again have to report poor results for the latest version of Cufflinks ($R^2=0.307$) with the version 0.9.3 still performing worse than the other three methods ($R^2=0.784$).

In the withing gene expression comparison (Table \ref{tab:relative}), we used two cutoffs for relevant transcripts. 
The first taking into account transcripts for which their gene has at least 10 reads in the ground truth (45662 transcripts) and the second considering only transcripts for which the gene has at least 100 reads (33757 transcripts).
BitSeq performs the best for the narrow range of transcripts with RSEM and MMSEQ having comparable results.
For the less stringent criteria, BitSeq still retains very good correlation with the ground truth while the performance of the other two methods deteriorates.
As we are using the same dataset, both versions of Cufflinks provide poor correlation when compared to other three methods.

\begin{table}[!h]
   \centering
\begin{tabular}{|c|c|c|c|c|c|}
\hline
  & BitSeq & Cufflinks & Cufflinks 0.9.3 & RSEM & MMSEQ \\
\hline 
above 10 reads & \textbf{0.951} & 0.205 & 0.739 & 0.876 & 0.888  \\
above 100 reads  & \textbf{0.964} & 0.176 & 0.787 & 0.945 & 0.948 \\
\hline
\end{tabular}

\bigskip
\caption{The $R^2$ correlation coefficient of estimated within-gene relative expression and ground truth.
The correlation was calculated for two groups, first one containing transcripts of genes with at least 10 reads and the second one containing transcripts of genes with at least 100 reads according to the ground truth.
}\label{tab:relative}
\end{table}

\subsection{Biological variance of RNA-seq data}

\begin{figure}[!tpb]
   \centering
   \includegraphics[width=0.7\textwidth]{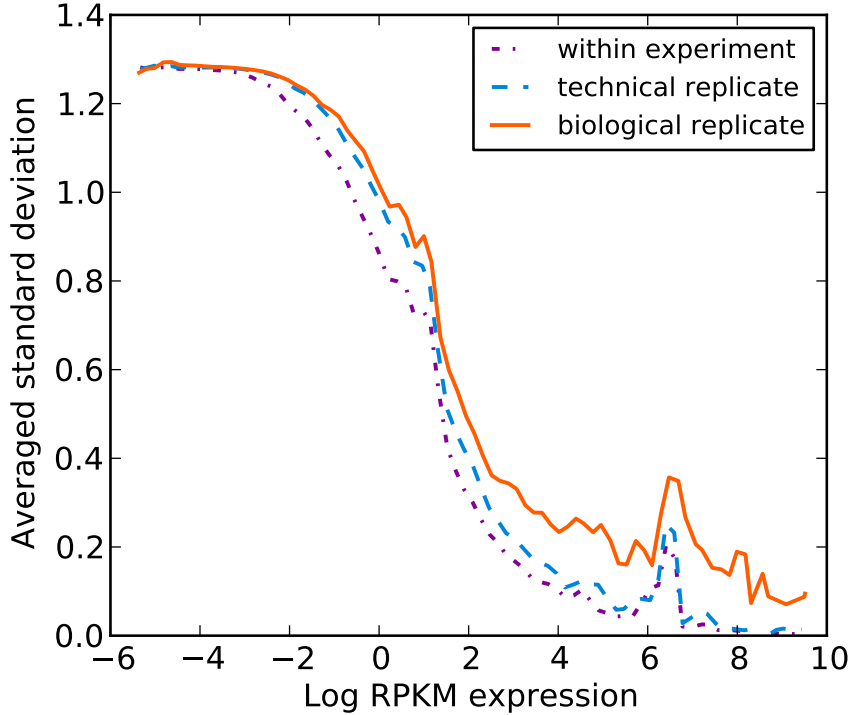}
\caption{Comparison of standard deviation of posterior samples within single dataset and combined datasets of technical replicates and biological replicates, with log RPKM expression on the x-axis and standard deviation of log RPKM expression on y-axis. 
The standard deviation is a sliding average over groups of transcripts with similar expression in order to highlight its dependents on the expression.
}\label{fig:variance}
\end{figure}

We used RNA-seq data from the microRNA target identification study \citep{Xu2010a} to test and compare the differential expression analysis method used in BitSeq Stage 2.
This dataset contains technical as well as biological replicates for each studied condition allowing assessment of the effects of biological variation.
Similarly to previous results \citep{Anders2010,Oshlack2010}, we observe significant biological variation within conditions.
Figure \ref{fig:variance} shows the standard deviation of transcript expression level posterior MCMC samples as a function of the mean expression level of the transcript. 
We compare the standard deviation for samples from within one experiment, between two technical replicates and between two biological replicates.
In order to calculate the standard deviation between replicates we took the squared root of variance which was estimated by computing mean square distance between samples.
Plotted values are averaged for a sliding window of similarly expressed transcripts. 
The MCMC sample variation captures the intrinsic estimation variance in the ``within-experiment'' case. 
The technical variance includes a contribution due to re-sequencing the same biological sample while the biological variance includes a contribution due to repeating the experiment.

We see that with higher expression the variation of the expression level estimation decreases as expected. 
At high expression levels the variance associated with technical replicates approaches the level of the within-experiment variance. 
On the other hand, the biological variance becomes relatively more significant in this regime. 
Without consideration of biological differences, high confidence of expression estimation of these transcripts will lead to false differential expression calls. 
It can also be observed that the within-experiment variance is a significant contribution to replicate variance (technical and biological) at lower expression levels. 
Therefore the intrinsic variance due to mapping ambiguity and limited read depth, as estimated by our MCMC expression estimation procedure, will provide useful information for assessing replicate variance in this low expression regime. 

\subsection{Assessing DE performance with simulated data}\label{sec:de_comparison2}

\begin{figure}[!h]
   \centering
   \includegraphics[width=0.80\textwidth]{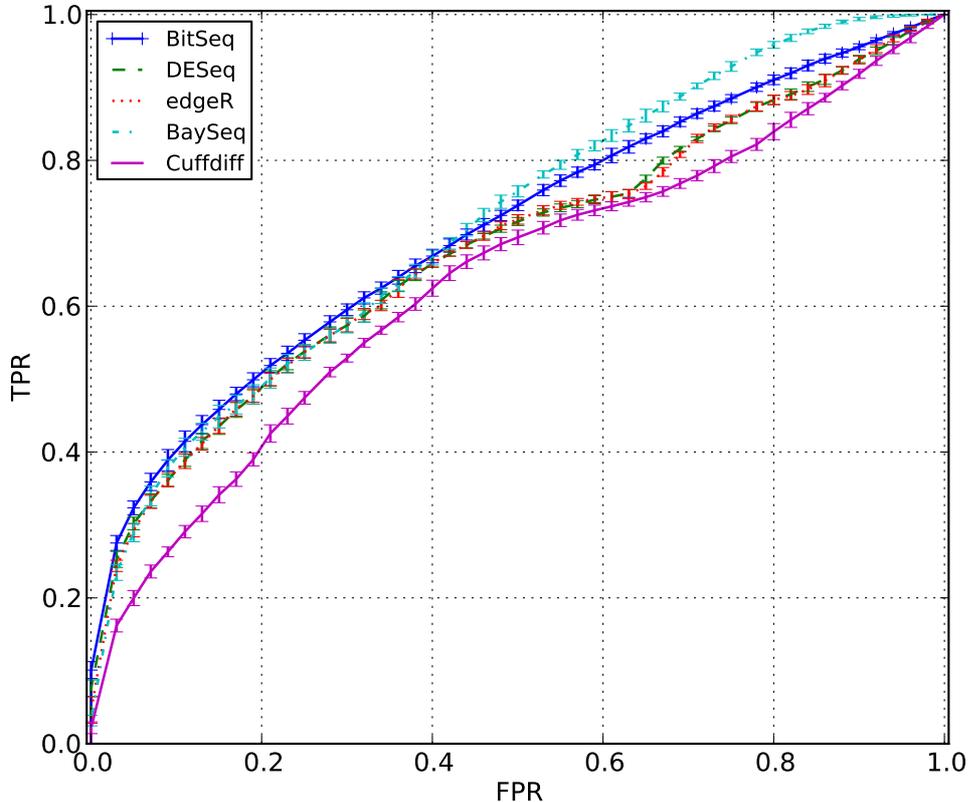}
   \caption{ROC curves averaged over 5 runs with standard deviation depicted by error bars. 
   The curve was calculated for transcripts with average of at least one read in the ground truth.
   The fold change was uniformly distributed in the interval $(1.5,3.5)$.
   }\label{fig:rocFull}
\end{figure}

\begin{figure}[!h]
   \centering
   \subfigcapskip-2.0ex
   \subfigcapmargin1ex
   \subfigure[]{
   \includegraphics[width=0.30\textwidth,trim=2mm 0mm 2mm 12mm,clip]{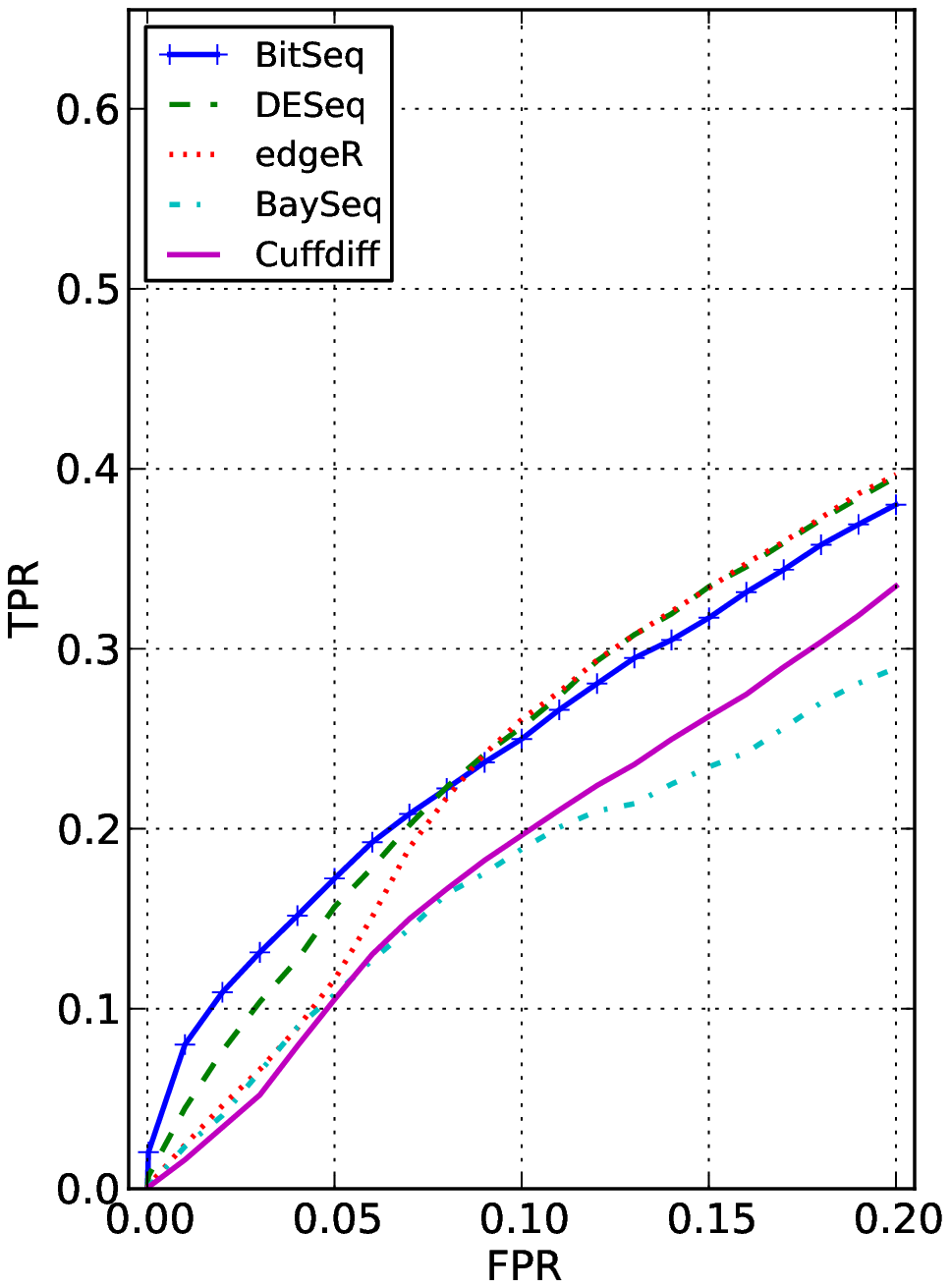}\label{fig:roc2-a}
   }
   \subfigure[]{
   \includegraphics[width=0.30\textwidth,trim=2mm 0mm 2mm 12mm,clip]{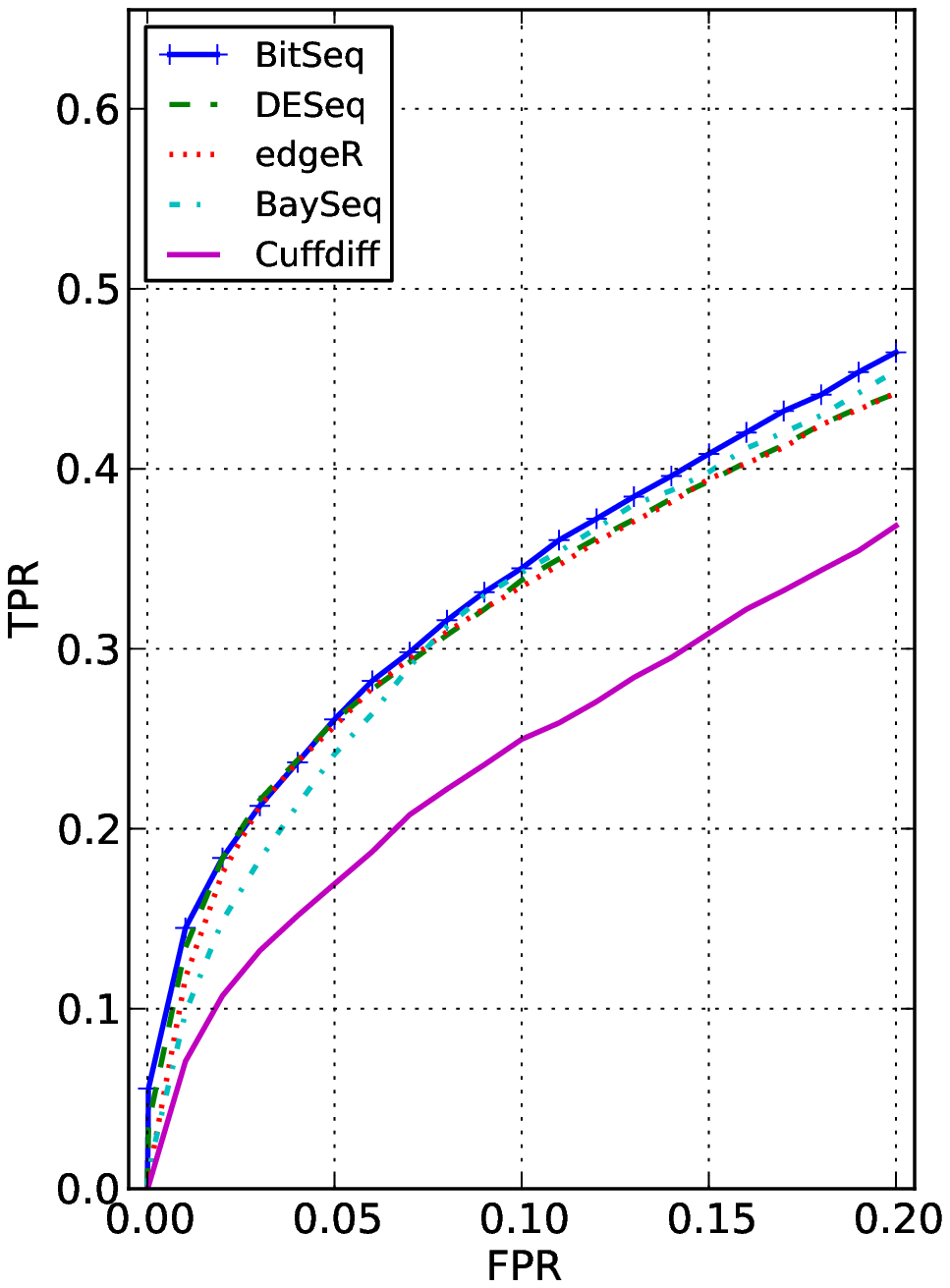}\label{fig:roc2-b}
   }
   \subfigure[]{
   \includegraphics[width=0.30\textwidth,trim=2mm 0mm 2mm 12mm,clip]{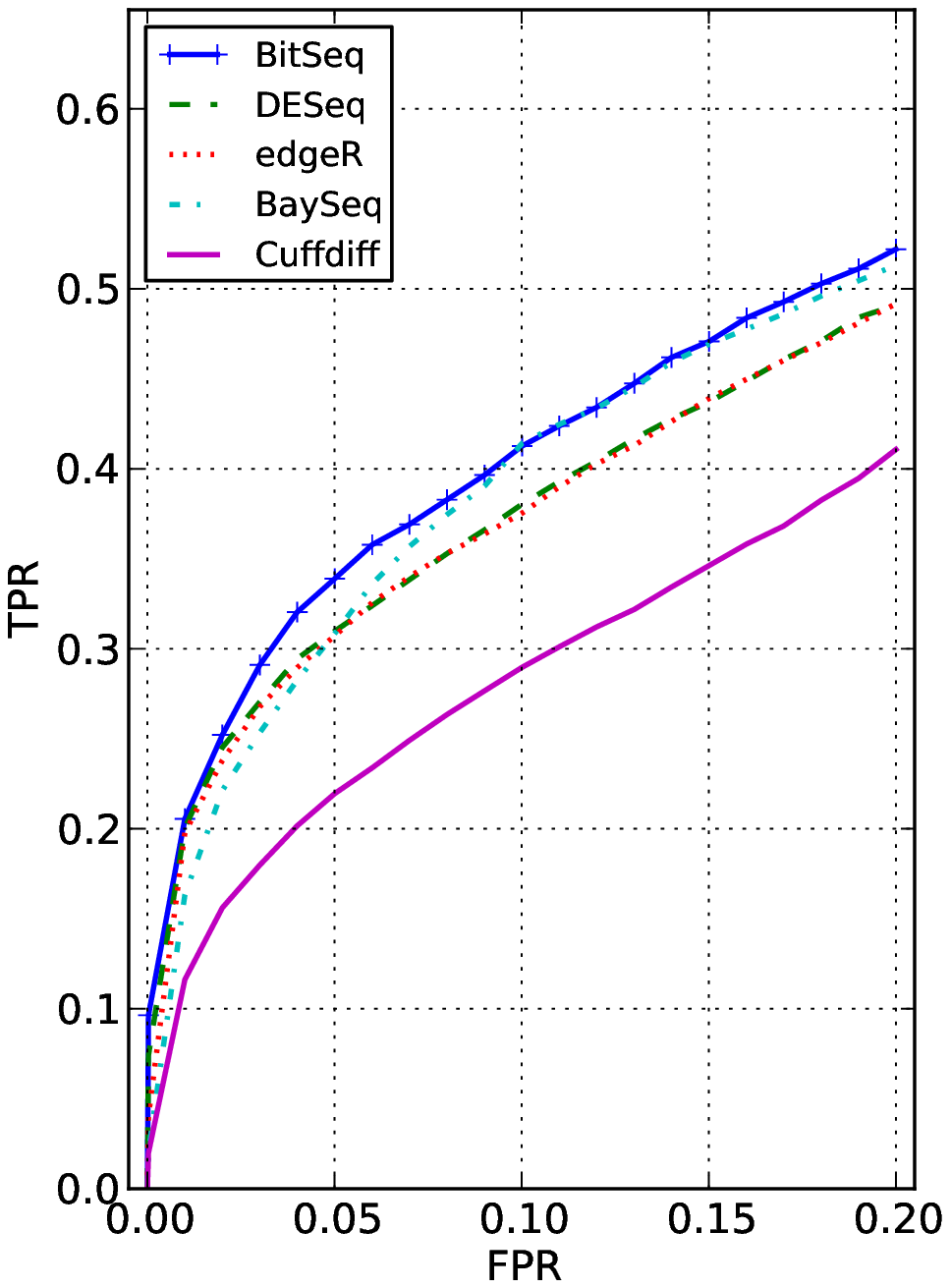}\label{fig:roc2-c}
   }
   \subfigure[]{
   \includegraphics[width=0.30\textwidth,trim=2mm 0mm 2mm 12mm,clip]{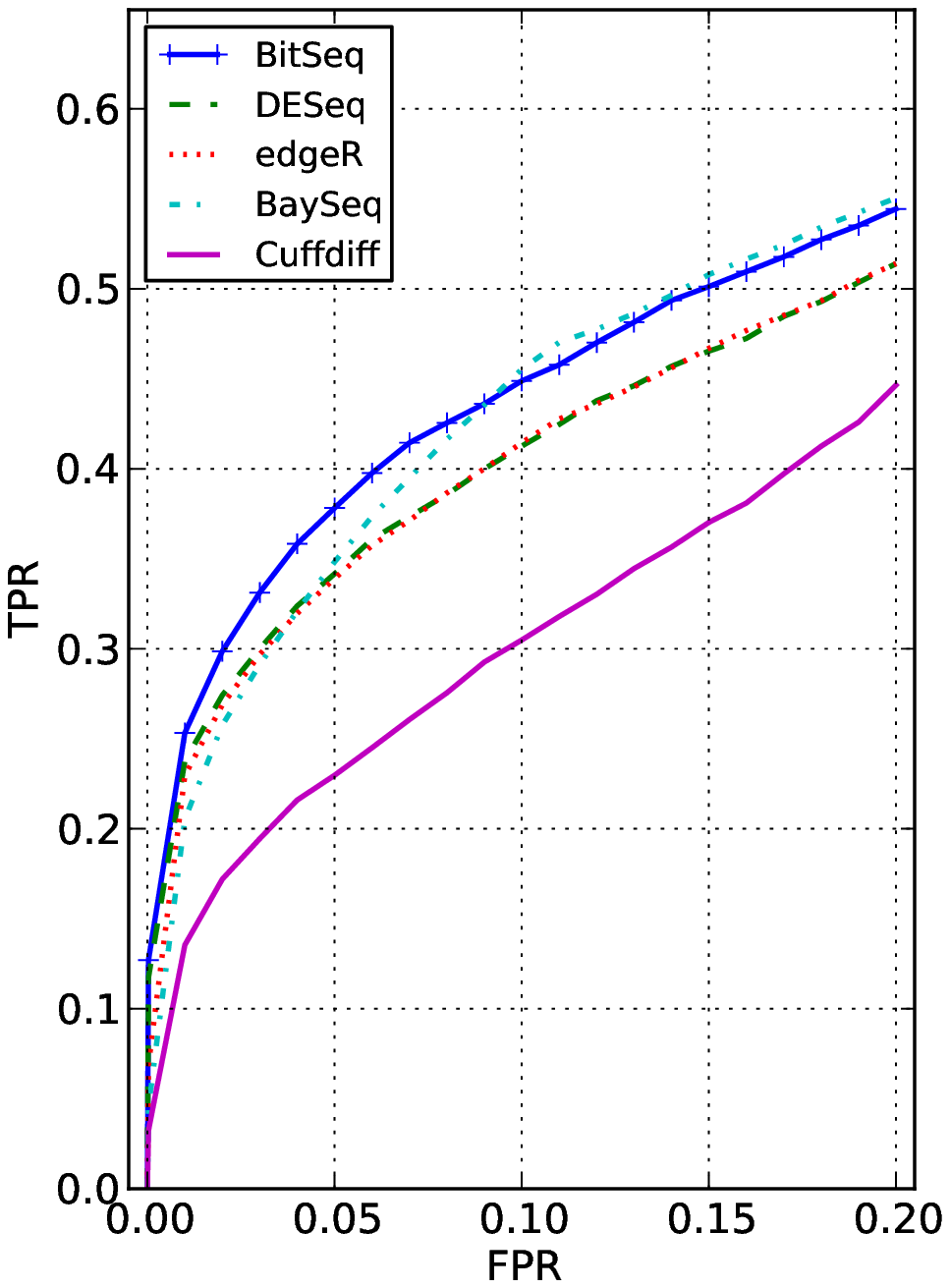}\label{fig:roc2-d}
   }
   \subfigure[]{
   \includegraphics[width=0.30\textwidth,trim=2mm 0mm 2mm 12mm,clip]{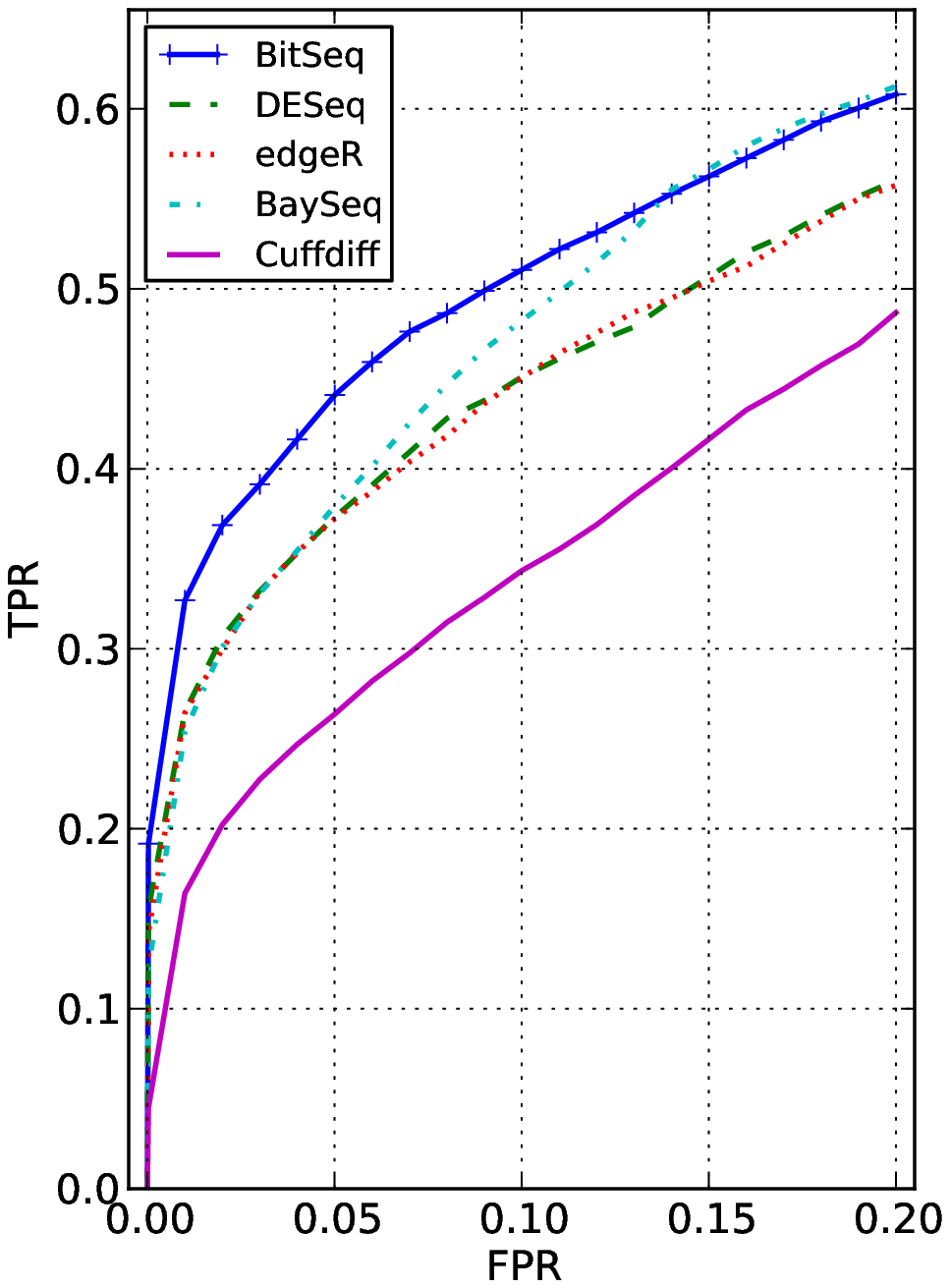}\label{fig:roc2-e}
   }
\caption{Differential expression analysis of simulated data with various levels of fold change.
The figures focus on the most relevant region with false positive rate above 0.2, and showing the y-axis up to true positive rate 0.65. 
The sub figures show data with fold change: 1.5 (a), 2.0 (b), 2.5 (c), 3.0 (d) and 5.0 (e). 
}\label{fig:roc}
\end{figure}

We carried out extensive assessment of DE analysis accuracy of BitSeq with comparison to other methods.
The Cuffdiff method from Cufflinks package \citep{Trapnell2010} is the only other method designed for transcript level DE analysis that uses replicates and accounts for biological variation.
We also included three popular methods which are primarily designed for gene level DE analysis (DESeq \citep{Anders2010}, edgeR \citep{Robinson2010b}, baySeq \citep{Hardcastle2010}), but given the lack of other options and their input being only the read count vectors, they could be considered for the transcript level analysis use case as well.
Using expression estimates obtained by BitSeq Stage 1, we converted the relative expression of fragments into read counts by simply multiplying it by the total number of aligned reads and used this as an input for the gene-level methods.
For each of these methods we used default parameter settings according to the packages' vignettes.

The Figure \ref{fig:rocFull} shows the same ROCs as Figure 6(a) in the main paper without the 0.2 cutoff. 
The evaluation is only for transcripts with at least one generated read on average with fold change being uniformly generated from the interval $(1.5,3.5)$.
In this figure, the error bars depict the standard deviation for the averaged curves showing consistent results trough the experiments.
We can see that BitSeq performs slightly better than the other methods with baySeq having higher true positive range in area with above 0.4 false positive range, however this area is not interesting from the application perspective. 

In the very last figure (\ref{fig:roc}), we compare the accuracy of these methods with respect to the fold change of differentially expressed transcripts.
We again restrict the figures to the area with false positive rate below 0.2 which in our opinion is the most important in terms of applicability.
Instead of using randomly selected fold change, all differentially expressed transcripts are either up-regulated or down-regulated by constant fold change.
The increase of fold change clearly improves the performance of the methods as we expected. 
BitSeq and baySeq have consistently better results than the other methods except for the lowest fold change $1.5$, in which baySeq has the lowest true positive rate and edgeR with DESeq outperform BitSeq in half of the spectrum.

In all of our DE experiments, Cuffdiff, despite being designed for transcript level analysis performs worse out of the 5 compared algorithm. 
This could be largely attributed to the expression estimation problem, however for DE analysis return to the older version (0.9.3) did not improve the results, possibly because of different DE model.
Our data also shows that for most parts, the DESeq and edgeR methods produce very similar results in terms of accuracy.
We have to note, that even though we tried to simulate the data in way to resemble real RNA-seq experiments, the data proved to be rather hard for all methods being compared.